\documentclass[article,nojss]{jss}
\usepackage{thumbpdf,lmodern} 
\graphicspath{{Figures/}}

\usepackage{amsmath}
\usepackage{bm}
\newcommand{\Ber}{\mathcal{B}er}
\newcommand{\ym}{\mathbf{y}} 
\newcommand{\zm}{\mathbf{z}} 
\newcommand{\partset}{{\cal C}}
\newcommand{\etav}{\boldsymbol{\eta}}
\newcommand{\parti}{{\cal P}}

\usepackage{fancyhdr}
\fancypagestyle{ack}{%
  \fancyhf{}

  \fancyfoot[L]{{\small This is a preprint of a chapter forthcoming in \emph{Handbook of Mixture Analysis},
    edited by Gilles Celeux, Sylvia Fr{\" u}hwirth-Schnatter, and Christian P. Robert.}}
}

\author{Bettina Gr{\" u}n\\Johannes Kepler University Linz}

\title{Model-based Clustering}

\Plainauthor{Bettina Gr{\" u}n}

\Abstract{
Mixture models extend the toolbox of clustering
methods available to the data analyst. They allow for an explicit
definition of the cluster shapes and structure within a probabilistic
framework and exploit estimation and inference techniques available
for statistical models in general.  In this chapter an introduction to
cluster analysis is provided, model-based clustering is related to
standard heuristic clustering methods and an overview on different
ways to specify the cluster model is given. Post-processing methods to
determine a suitable clustering, infer cluster distribution characteristics and
validate the cluster solution are discussed. The versatility of the
model-based clustering approach is illustrated by giving an overview
on the different areas of applications.}


\Address{
  Bettina Gr\"un\\
  Institut f\"ur Angewandte Statistik\\
  Johannes Kepler Universit{\"at} Linz\\
  Altenbergerstra{\ss }e 69\\
  4040 Linz, Austria\\
  E-mail: \email{Bettina.Gruen@jku.at}\\
  URL: \url{http://ifas.jku.at/gruen/}
}
\begin{document}

\thispagestyle{ack}
\section{Introduction}\label{sec:introduction}

Cluster analysis -- also known as unsupervised learning -- is used in
multivariate statistics to uncover latent groups suspected in the data
or to discover groups of homogeneous observations. The aim thus is often
defined as partitioning the data such that the groups are as
dissimilar as possible and that the observations within the same group
are as similar as possible. The groups composing the partition are
also referred to as clusters.

Cluster analysis can be used for different purposes. It can be
employed (1) as an exploratory tool to detect structure in
multivariate data sets such that the results allow to summarise and
represent the data in a simplified and shortened form, (2) to perform
vector quantisation and compress the data using suitable prototypes
and prototype assignments and (3) reveal a latent group structure
which corresponds to unobserved heterogeneity. A standard statistical
textbook on cluster analysis is for example
\citet{model-based:Everitt+Landau+Leese:2011}.

Clustering is often referred to as an ill-posed problem which aims at
revealing \emph{interesting} structures in the data or deriving a
\emph{useful} grouping of the observations. However, specifying what
is \emph{interesting} or \emph{useful} in a formal way is
challenging. This complicates the specification of suitable criteria
for selecting a clustering method or a final clustering
solution. \citet{model-based:Hennig:2015} also emphasises this
point. He argues that the definition of what the \emph{true} clusters
are depends on the context and the clustering aim. Thus there does not
exist a unique clustering solution given the data, but different
clustering aims imply different solutions and analysts should in
general be aware of the ambiguity inherent in cluster analysis and
thus transparently point out their clustering aims together with their
solutions obtained.

At the core of cluster analysis is the definition of what a cluster
is. This can be achieved by defining the characteristics of the
clusters which should emerge as output from the analysis. Often these
characteristics can only be informally defined and are not directly
useful to select a suitable clustering method.  In addition also some
notion on the total number of clusters suspected or the expected size
of clusters might be needed to characterise the cluster problem.
Furthermore, domain knowledge is important to decide on a suitable
solution, in the sense that the derived partition consists of
interpretable clusters which have practical relevance. However, domain
experts are often only able to assess the suitability of a solution
once they are confronted with a grouping but are unable to provide
clear characteristics of the desired clustering beforehand.

Model-based clustering can help in the application of cluster analysis
by requiring the analyst to formulate the probabilistic model which is
used to fit the data and thus making the aims and the cluster shapes
aimed for more explicit than what is generally the case if heuristic
clustering methods are used.  The use of mixture models for clustering
is also discussed in the general textbooks on mixture models by
\citet{model-based:McLachlan+Peel:2000b} and
\citet{model-based:Fruehwirth-Schnatter:2006}. In addition several
review articles on model-based clustering are available including
\citet{model-based:Stahl+Sallis:2012} and
\citet{model-based:McNicholas:2016} as well as the monograph on
mixture model-based classification by
\citet{model-based:McNicholas:2016a}.

In the following heuristic clustering methods are described and their
relation to Gaussian mixture modelling is elaborated. After discussing
the specification of the clustering problem, strategies for selecting
a suitable mixture model together with the appropriate clustering base
are pointed out in Section~\ref{sec:specifying-model}. The
post-processing methods given a fitted model are outlined in
Section~\ref{sec:post-processing} which allow to obtain an identified
model, derive a partition of the data, characterise the cluster
distributions, gain insights through suitable visualisations and
validate the clustering. Model-based clustering has been used in a
range of different applications where also methodological advances
have emerged from. Some important areas are discussed in detail in
Section~\ref{sec:applications}.

\subsection{Heuristic clustering}\label{sec:heuristic-clustering}

Heuristic clustering methods are based on the definition of
similarities or dissimilarities between observations and groups of
observations. These are used as input to the different cluster
algorithms proposed. In general two main clustering strategies are
distinguished:   hierarchical clustering and  partitioning
clustering.

In hierarchical clustering a nested sequence of partitions is
constructed. This is accomplished either in a bottom-up
(\emph{agglomerative}) or a top-down (\emph{divisive}) approach. In
bottom-up or agglomerative clustering, each observation starts in its
own cluster and in each step two clusters are merged until all
observations are contained in one single cluster. By contrast top-down
or divisive clustering starts with a single cluster and in each step
splits one cluster into two. A greedy search strategy
is employed where in each single step the optimal one-step ahead decision is
made. However, this does not imply that any of the intermediate
cluster solutions obtained is optimal.

In order to obtain an optimal partition of $n$ observations  $\ym=(y_1, \ldots, y_n)$ for a given number of
clusters $G$,  a partitioning cluster algorithm needs to be used. The
$k$-means algorithm aims at finding the partition $\parti$ which minimises
the within-cluster sum-of-squares, 
\begin{align*}
  \arg\min_{{\parti}} \sum_{g=1}^G \sum_{i \in {\partset}_g} \sum_{j \in {\partset}_g} \| {y}_i - {y}_j\|^2,
\end{align*}
where ${\partset}_g$ is the set of observations assigned to group $g$ by the
partition ${\parti}$ and $\| \cdot \|$ denotes the Euclidean distance. The
solution to this objective function can also be obtained by solving
the following optimisation problem
\begin{align*}
  \arg\min_{{\parti}} \sum_{g=1}^G \sum_{i \in {\partset}_g} \| {y}_i - {\bar{y}_g}\|^2 ,
\end{align*}
where the cluster centroid ${\mu}_g$ is equal to $\bar{y}_g$ which is
given by
\begin{align*}
  \bar{y}_g &= \frac{1}{n_g} \sum_{i \in {\partset}_g} {y}_i,
\end{align*}
with $n_g$ the number of observations in group $g$. Note that if the
partition is restricted to contain only non-empty elements, $G$ is
necessarily finite for a finite sample of size $n$, but if the
partition $\parti$ may also contain empty groups $G = \infty$ is also
possible. This observation similarly holds also for the mixture models
subsequently considered. For mixture models finite or infinite values
for $G$ might be used to represent the theoretic data-generating
process. However, the induced partitions will consist of a finite
number of groups for finite samples. Also the number of components
with observations assigned will be finite for finite samples.

Extension of the $k$-means algorithm to alternative dissimilarity
measures than the squared Euclidean distance have been proposed
leading to $k$-centroid cluster analysis
\citep{model-based:Leisch:2006}. These variants can also be used for
multivariate data where variables are collected on different scale
levels and which require specific dissimilarity measures. For example
for asymmetric binary data the Jaccard distance or Jaccard coefficient
\citep{model-based:Jaccard:1912} has been proposed as suitable
dissimilarity measure because it disregards joint zeros \citep[see,
for example,][p.~26]{model-based:Kaufman+Rousseeuw:1990}.

\subsection{From $k$-means to Gaussian mixture modelling}\label{sec:from-k-means}

If finite mixtures of multivariate Gaussian distributions are used for
model-based clustering, a probabilistic distribution is specified which
is used as a data-generating process for the observed data. In
particular it is assumed that the data in each group or cluster is
generated from a multivariate Gaussian distribution and the combined
data stems from a convex combination of multivariate Gaussian
distributions. This distribution used in Gaussian mixture modelling is
given by
\begin{align*}
  \sum_{g=1}^G \eta_g \phi({y} | {\mu}_g, \Sigma_g),
\end{align*}
where $\phi(\cdot | \mu, \Sigma)$ is the pdf of the multivariate
Gaussian distribution with mean $\mu$ and covariance matrix $\Sigma$,
corresponding to the cluster distribution, and $\eta_g$ are the
cluster sizes with $ \eta_g \geq 0, \sum_{g=1}^G \eta_g = 1.$ The
parameters $\theta$ in this model are the cluster sizes $\eta_g$, and
the cluster-specific parameters consisting of the cluster means
$\mu_g$ and the cluster covariance matrices $\Sigma_g$ for
$g=1,\ldots,G$.

This model class can be seen as a generalisation of $k$-means
clustering. Define  $\zm=(z_1, \ldots, z_n)$, where $z_i \in \{1,\ldots,G\}$ is the cluster membership of
observation $i$. 
The $k$-means clustering solution can also be obtained by
maximising the classification likelihood $p(\ym|\zm,\theta)$  of a finite mixture of
multivariate Gaussian distributions with identical isotropic covariance matrices $\Sigma_g= \sigma^2 I$, where  $I$ is the identity matrix,
 and equal weights $\eta_g=1/G$ with respect to the
mixture parameters $\theta$  and the cluster memberships $\zm$: 
\begin{align*}
  \sum_{i=1}^n \frac{1}{G} \phi(y_i | \mu_{z_i}, \sigma^2 I).
\end{align*}
This implies that quite specific cluster shapes are implicitly imposed
when using $k$-means clustering, namely that the clusters are
spherical with equal variability in each dimension. In such a
situation it is important to define a suitable scaling for each of the
dimensions because the $k$-means solution treats the variability in
each of the dimensions equally and the obtained solution therefore is
not invariant with respect to the scaling of the variables. A
data-driven approach to achieve this is often to standardise the
data. This, however, is problematic if (1) extreme, outlying
observations are present in the data and (2) the cluster structure is
not equally strong in all dimensions leading to different
between-cluster dissimilarities in the different dimensions. The
dissimilarities in dimensions containing strong cluster structure are
attenuated by the standardisation and thus standardisation
deteriorates the obtained cluster solution.  In addition the
application of $k$-means favours clusters of the same size and
volume. Thus making this relationship between Gaussian mixture models
and $k$-means clustering explicit allows to gain insights into the
implicit assumptions imposed when $k$-means clustering is used. The
notion that heuristic clustering methods impose less assumptions than
model-based clustering is thus clearly in error. Rather users are
often less aware of the assumptions implicitly imposed when using
heuristic methods.

Extending the $k$-means approach to allow for arbitrary covariance
matrices in the clusters leads to model-based clustering using finite
mixtures where the classification likelihood $p(\ym|\zm,\theta)$
instead of the observed likelihood $p(\ym|\theta)$ is maximised
\citep{model-based:Scott+Symons:1971, model-based:Symons:1981,
  model-based:McLachlan:1982, model-based:Celeux+Govaert:1992}. In
this case the parameters of the mixture model as well as the cluster
memberships are estimated, implying that the number of quantities to
be estimated grows with the number of observations.  Under these
conditions maximum likelihood estimates have been shown to be
asymptotically biased \citep{model-based:Bryant+Williamson:1978}.

The classification maximum likelihood approach can be implemented in
two different ways: either excluding the cluster sizes
$\etav=(\eta_1, \ldots, \eta_G)$ from the likelihood or including
$\etav$ in the likelihood. The first case corresponds to implicitly
assuming that they are all equal, while the second case can be seen as
adding a penalty term to the classification likelihood. Maximising the
classification likelihood instead of the observed likelihood has the
advantage that the derived solution potentially is better suited for
the clustering context and that iterative methods for model estimation
such as the expectation-maximisation (EM) algorithm converge
faster. \citet{model-based:Biernacki+Celeux+Govaert:2003} thus
consider to initialise the ordinary EM algorithm by first using
several runs of the so-called classification EM, which implements an
EM algorithm for maximising the classification likelihood, and
selecting the best solution detected.

Applications of Gaussian mixture modelling are widespread and often
this model class is used as the basic model for clustering metric
multivariate data if $k$-means clustering is not flexible
enough. Recent exemplary applications of Gaussian mixture modelling are, among many others, 
\citet{model-based:Kim+Yun+Park:2014} in hydrology who cluster a
multivariate data set of hydrochemical measurements from groundwater
samples to separate anthropogenic and natural groundwater groups.
\citet{model-based:Perera+Mo:2016} used Gaussian mixture models in
ocean engineering to understand marine engine operating regions as
part of the ship energy efficiency management plan. In environmental
science,  \citet{model-based:Skakun+Franch+Vermote:2017} used Gaussian
mixture models to determine a data-driven classification method to
distinguish winter crop from spring and summer crop.

Compared to $k$-means clustering Gaussian mixture modelling has
several advantages. The model explicitly allows for clusters of
different siz and clusters of different volume. In addition, the
clusters are independent of the scaling used for the variables (except
for potential numerical issues).  This flexibility comes with several
drawbacks. The likelihood is unbounded for the general Gaussian
mixture model and spurious solutions might emerge. Assuming a
one-to-one mapping between clusters and components in the mixture
model might lead to counter-intuitive clustering solutions, because
(a) several components form a single mode and (b) observations are not
assigned to the cluster where the component mean is closest in
Euclidean space due to different component-specific covariance
matrices.  Finally, the fitted model might rather correspond to a
semi-parametric estimation of the data distribution, than represent a
useful clustering model.
  
Thus specification of a suitable model (see
Subsection~\ref{sec:specifying-model}) and the use of appropriate
post-processing techniques, discussed in
Subsection~\ref{sec:post-processing}), are essential to reach a
cluster solution that is useful and meets the stated clustering aims.

\subsection{Specifying the clustering problem}\label{sec:spec-clust-probl}

\cite{model-based:Hennig+Liao:2013} claim that ``there are no unique
`true' or `best' clusters in a data set'', rather it needs to be
defined by the user who applies clustering methods what a cluster
is. In general this consists of specifying the characteristics of a
cluster regarding size and shape and how clusters are assumed to
differ. This constitutes essential information in order to assess
which observations form a cluster and which belong to different
clusters. These decisions need to be made regardless of whether a
model-based approach or a heuristic approach to clustering is
employed. However, using a model-based approach makes these decisions
in general more explicit. The specified model clearly indicates what
cluster distributions are considered. Furthermore, in a model-based
approach model selection and evaluation are based on statistical
inference methods. This allows to recast the problem of choosing a
suitable number of clusters as a statistical model selection problem
and adds the possibility to rigorously assess uncertainty.

Different notions of what defines a cluster have been proposed and
common examples of such cluster characteristics are described in the
following consisting of compactness, density-based levels,
connectedness and functional similarity.  For illustration,
2-dimensional scatter plots of data where a clear cluster structure
regarding these notions is present are used.

\emph{Compactness.} A cluster is characterised by points being close
  to each other. Separation between observations indicates that they
  stem from different clusters. In this case the cluster centroid is a
  useful prototype for all observations in the cluster.
  Often the notion of compactness is used to derive a cluster solution
  assuming that all clusters have similar levels of
  compactness. This implies that all clusters have a comparable
  volume and that the cluster centroids equally well represent
  observations in their clusters.  The $k$-means algorithm minimises the
  within-cluster sum-of-squares which can be seen as a measure of
  compactness and thus explicitly tries to address this clustering
  notion. In hierarchical clustering some linkage methods, i.e.,
  distance definitions between groups of observations, also lead to
  solutions with high compactness, such as complete linkage.

\emph{Density-based levels.}  Areas in the observation space where
  observations frequently occur are referred to as density clusters.
  This cluster concept in general is used for continuous data. In
  contrast to the compactness notion clusters might have different
  volumes and arbitrary shapes, i.e., also non-convex shapes, are
  possible.  However, this also implies that the centroids -- a concept
  which might even not be well defined for non-convex clusters -- do
  not equally well represent their assigned observations. Furthermore, new
  observations might be hard to assign to a cluster if they occur in
  regions not identified as cluster regions, i.e., assigning any (new)
  observations to one of the clusters might not be straightforward and
  unambiguous.   Cluster methods which are based on an estimate of the data density
  and determine connected components in a density level set try to
  explicitly address this cluster notion
  \citep{model-based:Azzalini+Menardi:2014, model-based:Scrucca:2016}.

\emph{Connectedness.} A cluster is defined by a friends-of-friends
  strategy. Observations are assigned to the same cluster if they are
  close to each other or if they are linked by other observations
  also assigned to the cluster. This implies that arbitrary cluster shapes
  are possible. However, no further structure than provided by the
  data itself is imposed indicating that solutions might be quite
  data-dependent and variable. Also characterising a cluster through
  a centroid or a prototype might be impossible.
  Some linkage methods used in hierarchical clustering favour this kind
  of solutions, such as single-linkage where a \emph{chaining
    phenomenon} occurs.

\emph{Functional similarity.} Observations are assigned to the same
  cluster if they share a common functional relationship between the
  variables in the different dimensions. For example, one variable
  might be seen as the dependent and the others as independent ones
  leading to a regression setting. Functional similarity then implies
  that the regression coefficients are similar within clusters, but
  differ across clusters. Network data represent another data setting
  where functional similarity might serve as notion for constructing
  clusters. In this case groups are for example formed by joining
  observations which have a similar linking behaviour to other
  observations.

\begin{figure}[t!]
  \centering
\includegraphics[width=\textwidth]{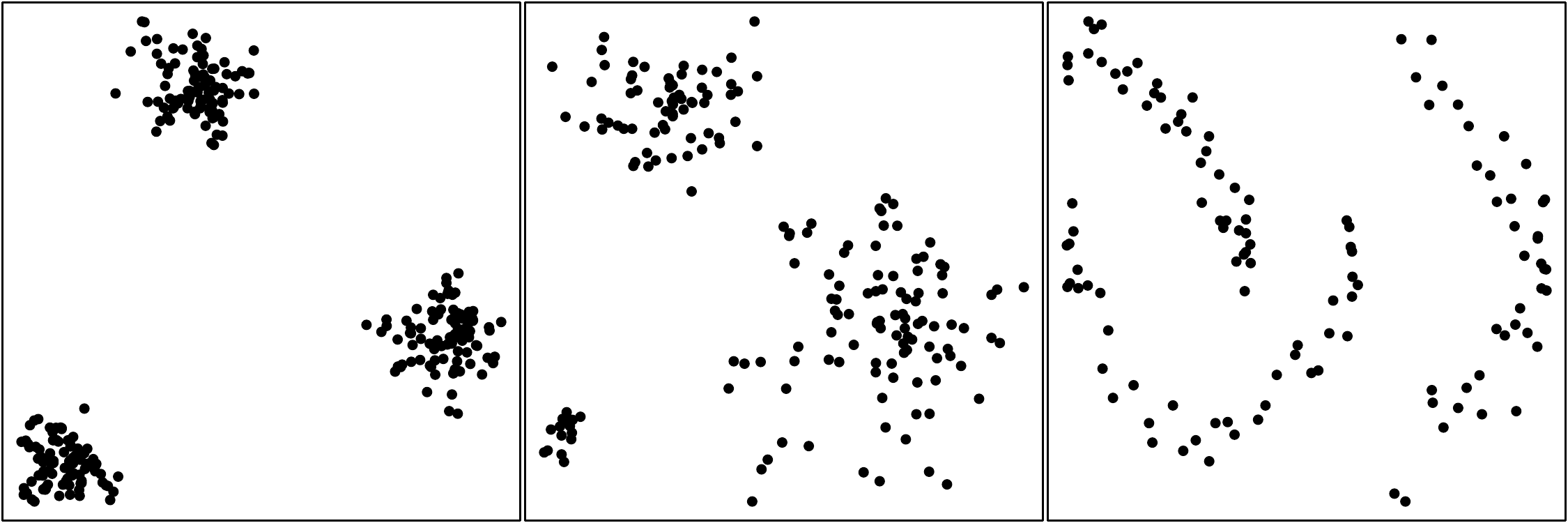}
\caption{Illustration of clustering concepts:
  (a) compact clusters (on the left);
  (b) density-based clusters (in the middle);
  (c) connected clusters or clusters sharing a functional relationship (on the right).}
  \label{fig:clustering-concepts}
\end{figure}

\begin{figure}[t!]
  \centering
\includegraphics[width=\textwidth]{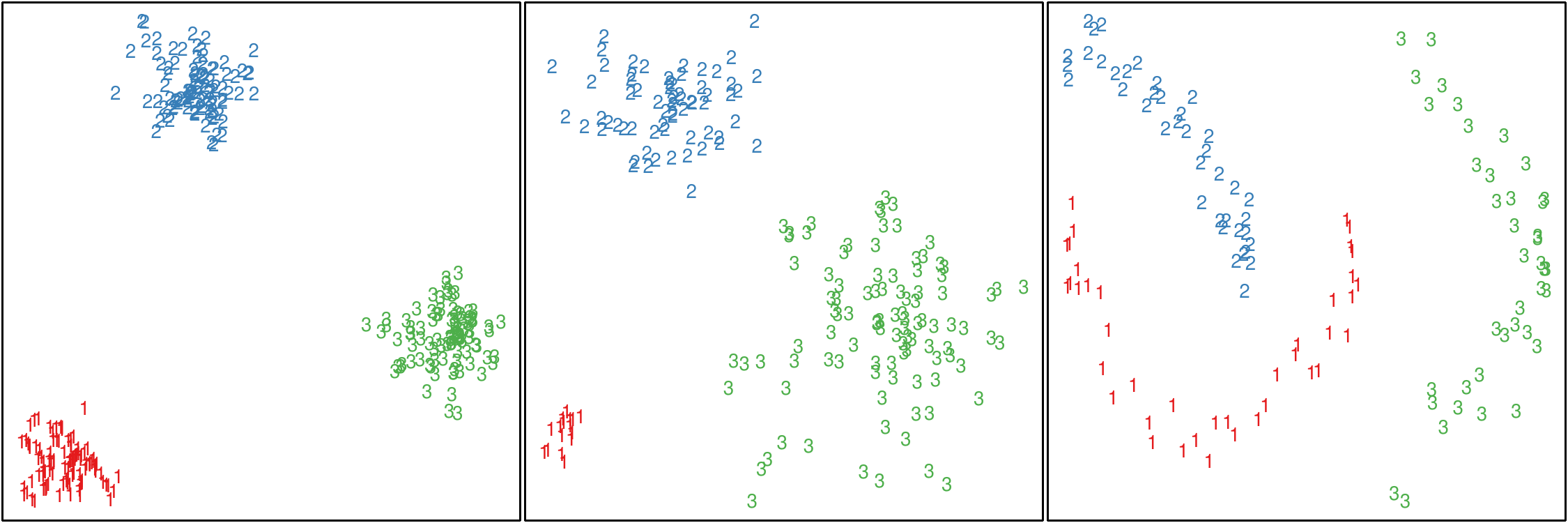}
\caption{Illustration of clustering concepts including the true
  cluster membership indicators indicated by colour and using numbers
  for the data points: (a) compact clusters (on the left); (b)
  density-based clusters (in the middle); (c) connected clusters or
  clusters sharing a functional relationship (on the right).}
  \label{fig:clustering-concepts-with-cl}
\end{figure}

Figure~\ref{fig:clustering-concepts} illustrates these different
clustering concepts using three data sets containing two-dimensional
metric data. The same data is also shown in
Figure~\ref{fig:clustering-concepts-with-cl} where the information on
the cluster membership of each observation is also included using
different colours and numbers for the data points. The scatter plot on
the left visualises a data set where the clusters are compact, i.e.,
they have a similar level of compactness and shape and are very well
separated. This is the easiest case for clustering where most of the
clustering methods employed should be able to detect the correct
cluster structure.  The scatter plot in the middle shows a data set
where clusters correspond to high density levels and can still be
represented by a cluster centroid, but differ in their level of
compactness. In contrast to $k$-means, model-based clustering allows
to account for these differences in compactness and to better extract
the true cluster structure.  The data set given in the scatter plot on
the right illustrates the case where clusters can easily be identified
using the connectedness concept. In addition these connected clusters
might be identifiable by imposing cluster-specific functional
relationships between the values on the $x$- and the $y$-axis and also
by assuming that cluster sizes vary with the value of $x$.  If
heuristic clustering methods are used for the data set on the right,
these need to be able to detect clusters based on connectedness.

If $k$-means is used to cluster the three data sets, good performance
is expected for the first two which are based on the compactness and
density-based concepts for clusters. $k$-means is expected to perform
badly on the third data set and not to be able to detect the true
cluster structure. The potential performance of $k$-means clustering on
these data sets is investigated in
Section~\ref{sec:visu-clust-solut}. The quality of the true clustering
solutions is visualised in Figure~\ref{fig:silhouettes} using a
silhouette plot based on the Euclidean distance.  This indicates how
$k$-means clustering might potentially perform. In addition, a
silhouette-type plot is also used to illustrate how model-based
clustering methods might perform on these three artificial data sets.
The conditional probabilities of cluster memberships are determined
based on suitable mixture models fitted using maximum likelihood
estimation. These are then split by the true cluster memberships of
the observations and visualised in Figure~\ref{fig:posteriors}.

The application of standard model-based clustering methods in general
does not ensure that compact, high density or connected clusters or
clusters with distinctly different functional relationships are
obtained and heuristic methods might be easier to tune to address any
of these notions. However, this also implies that the heuristic
methods are more rigid in the solutions detected and might be more
likely to come up with a specific  clustering solution regardless of the
inherent structure in the data. Thus solutions obtained using
model-based clustering techniques which are agnostic of these cluster
requirements can be assessed regarding these characteristics to verify
if such a structure might inherently be present in the data. This also
emphasises the need for specifying a suitable mixture model as well as
applying appropriate post-processing methods to ensure that the
outcome of model-based clustering is a useful clustering solution.

\section{Specifying the model}\label{sec:specifying-model}

For model-based clustering, finite as well as infinite mixture models
have been used to fit a suitable distribution to the data and, in a
subsequent step, to infer the cluster memberships from the fitted
distributions and, potentially, also gain insights about typical
characteristics of the cluster distributions.
\citet{model-based:Fraley+Raftery:2002} point out that model-based
clustering embeds the cluster problem in a probabilistic
framework. The statistical framework allows to employ statistical
inference methods for obtaining a suitable clustering of the data.

The starting point for model-based clustering is to define the
distribution of a cluster and to decide how the cluster sizes are
distributed.  The definition of the cluster distribution can also be
seen as the specification of the clustering kernel
\citep[see][]{model-based:Fruehwirth-Schnatter:2011}. Depending on the
available data and its characteristics, different kernels are
suitable. In addition the purpose of the cluster analysis and
characteristics of the intended cluster solution also guide the choice
of the clustering kernel. The requirement that a cluster distribution
needs to be specified makes model-based clustering in general more
transparent with respect to the targeted clustering solutions, while
in applications of heuristic methods the characteristics of the
targeted clustering solutions often remain implicit.

Heuristic clustering is based on notions of similarity and
dissimilarity between observations and groups of
observations. Model-based methods also use a notion of similarity
between observations. In the case of model-based methods observations
in the same cluster share the characteristic that they are generated
from the same cluster distribution.  Observations are ``similar'' if
they are generated by the same cluster distribution. The dissimilarity
between observations in different clusters is influenced by the
differences allowed for the cluster distributions.

If the same parametric distribution is assumed for each of the
clusters, differences between cluster distributions are determined by
the specific parameter values only.  In the case of mixtures of
multivariate Gaussian distributions for example often a centroid-based
approach is pursued. The cluster means are assumed to differ and
characterise the cluster distributions while the cluster covariance
matrices rather represent nuisance parameters. The cluster covariance
matrices might differ over clusters, but potentially could also be
similar. If the covariance matrices are fixed to be identical over
clusters, the dissimilarity between clusters is solely based on the
cluster centroids. The conditional probabilities of cluster membership
of an observation are then based on the Mahalanobis distance between
the observation and the cluster centroid weighted by the cluster
sizes. 

\subsection{Components corresponding to clusters}\label{sec:comp-corr-clust}

If mixture models are used for model-based clustering, the standard
assumption is that each of the components in the mixture model
corresponds to a cluster in the clustering solution. This implies that
the component distribution specified in the mixture model is also the
assumed cluster distribution.

For the specification of the mixture model it needs to be decided
first if the clusters are expected to all follow a distribution from
the same family of parametric distributions and that they differ only
in the values of the parameter vectors or if different parametric
distributions are assumed for the different clusters. The later might
only be possible in cases where a clear notion of the latent groups to
be detected is available. For instance,  if the presence of two groups
corresponding to healthy and ill persons is assumed and also some
prior knowledge is available that the distributions of these two
groups are structurally different.  Otherwise -- which is the usual
case -- the model specification is a priori the same for all of the
components, i.e., the same (parametric) model is imposed and
components differ only in the specific values of the parameters.

The choice of distribution for the components is governed by two
aspects: (1) the data structure and (2) the suspected cluster
distributions. In particular the data structure can easily be verified
and has led to a standard set of mixture models to be considered for
certain kinds of data, e.g., multivariate continuous data,
multivariate categorical data, multivariate mixed data or longitudinal
data.

\paragraph{Multivariate continuous data.}
For model-based clustering of multivariate continuous data of dimension $d$ the
standard model is the finite mixture model of multivariate Gaussian
distributions given by
\begin{align*}
  y_i &\sim \sum_{g=1}^G \eta_g \phi(y_i | \mu_g, \Sigma_g).
\end{align*}
The advantages of Gaussian mixture modelling are that estimation
methods are well established and that the component distributions are
thoroughly understood and thus interpretation of results is
facilitated. Drawbacks are that the likelihood is unbounded for
arbitrary covariance matrices $\Sigma_g$ because of singular solutions
leading to spurious results. Results are also strongly susceptible to
extreme observations because of the light tails of the Gaussian
distribution.  When clustering data,  it is often essential that the
means of the cluster distributions are different and can be used as
centroids to represent the clusters. However, this is not necessarily
achieved when using this standard model class because there are no
constraints -- neither implicitly nor explicitly -- imposed which
would ensure that the fitted mixture model has these characteristics.

If a centroid-based approach to clustering is pursued, the
covariance matrices of the components in the mixture model
are seen as nuisance parameters which need to be suitably
parameterised to allow for identification of the clusters. In
particular for high-dimensional data the specification of the
covariance matrices is crucial because they add a substantial
number of additional parameters to the model as the number of
parameters of a single, unrestricted covariance matrix is
$d (d+1) / 2$ for data of dimension $d$.

To achieve a parsimonious parameterisation,  different variants to
impose restrictions on the covariance matrices based on the spectral
decomposition of the covariance matrix into shape, volume and
orientation have been proposed in
\citet{model-based:Banfield+Raftery:1993} and are also discussed in
\cite{model-based:Celeux+Govaert:1995} and
\cite{model-based:Fraley+Raftery:2002}. The decomposition of the
covariance matrix of the $g$-th component is given by
\begin{align*}
 \Sigma_g &= \lambda_g D_g A_g D_g ^\top,
\end{align*} 
where the positive scalar $\lambda_g$ corresponds to the volume, the $d\times d$ matrix $D_g$ to the
orientation and the $d$-dimensional diagonal  matrix $A_g$ to the shape. More parsimonious
parameterisations can be achieved by restricting the components of the
decomposition to be equal over components and by imposing certain
values, e.g., assuming $D_g$ to be the identity matrix.

An alternative approach for a parsimonious parameterisation of the
covariance matrices is to assume a latent structure leading for
example to factor analysers as models for the covariance matrices
\citep{model-based:McLachlan+Peel:2000} or the unified latent
Gaussian model approach which encompasses factor analysers and
probabilistic principal component analysers as special cases
\citep{model-based:McNicholas+Murphy:2008}.

The use of mixtures of Gaussian distributions for clustering data
imposes a prototypical shape on the clusters which implies
symmetric and light tailed distributions. However, often in
applications clusters are assumed to have different shapes, e.g., data
clusters could exhibit skewed shapes and contain outlying
observations. In order to account for skewness the extension to
multivariate skew-normal distributions
\citep{model-based:Azzalini+Dalla-Valle:1996} has been considered and
for a more robust method the $t$-distribution is used for the
components. For an application in the mixture context see
\cite{model-based:Fruehwirth-Schnatter+Pyne:2010},
\cite{model-based:Lee+McLachlan:2013} and
\cite{model-based:Lee+McLachlan:2014}. Further approaches considered
for example the use of shifted asymmetric Laplace distributions
\citep{model-based:Franczak+Browne+McNicholas:2014} or multivariate
normal inverse Gaussian distributions
\citep{model-based:Hagan+Murphy+Gormley:2016}.

\paragraph{Multivariate categorical data.}

In the context of multivariate binary data the latent class model has
been developed as a useful tool for clustering
\citep{model-based:Goodman:1974}. This model can also be easily
extended to the case of categorical data. Latent class models aim at
explaining the dependency structure present in the multivariate data
by introducing a discrete latent variable, i.e., the cluster
membership, such that conditional on the latent variable the
observations in the different dimensions are independent.  This
assumption is also referred to as \emph{conditional local
  independence}.  The latent class model for multivariate binary data
is given by
\begin{align*}
y_i  \sim   \sum_{g=1}^G \eta_g \left[\prod_{j=1}^d
  f_{\Ber}(y_{ij} | \pi^j_{g})\right],
\end{align*}
where $f_{\Ber}(\cdot | \pi)$ is the density of the Bernoulli
distribution with the success parameter $\pi$ and $\pi^j_g$ is the
success probability for the $j$-th variable and the $g$-th component.

\paragraph{Multivariate mixed data.}

The conditional local independence assumption can also be used for
modelling mixed data. For groups of variables where a joint
distribution allowing for dependencies might be hard to specify for
the components,  a product of the univariate distributions is used
\citep{model-based:Hunt+Jorgensen:1999}. Thus as long as a suitable
parametric distribution is available in the univariate case a mixture
model can be specified and fitted.

Alternatively an approach has been proposed which uses the following
model for the components: the data in each cluster is assumed to be
generated based on a latent variable which follows a multivariate
Gaussian distribution where an arbitrary covariance matrix can be
used. The latent variable is then mapped to the observed measurement
scale, e.g., a binary or ordinal variable
\citep{model-based:Cai+Song+Lam:2011,
  model-based:Browne+McNicholas:2012,
  model-based:Gollini+Murphy:2014}.

\paragraph{Multivariate data with special structures.}

Special structures in multivariate data sets occur because of the
variables having different roles or because of specific constraints
restricting the values of the variables which can jointly be observed.
If one variable represents a dependent variable and the others
explanatory variables this leads to a regression setting. Multivariate
data occurring only on a restricted support than the one induced by
taking the product of the support of each of the univariate variables
require to take this into account in model specification. Graph or
network data also have a specific multivariate data structure.

In a regression setting one variable is identified as the dependent
variable $y$ and all other variables act as potential independent
variables $x$. The mixture model is then given by
\begin{align*}
  y_i | x_i &\sim \sum_{g=1}^G \eta_g f(y_i | x_i, \theta_g),
\end{align*}
where $f(\cdot)$ represents a regression model and $\theta_g$ contains
the regression parameters of component $g$. Clearly any regression
model could be used for the components and identified with a
cluster. The selection of the regression models for the components
needs to ensure that any other inherent structure in the multivariate
data set is captured. For example, repeated observations might be taken
into account by including random effects.

Identifiability might be an issue in the context of mixtures of
regressions \citep{model-based:Follmann+Lambert:1991,
  model-based:Hennig:2000, model-based:Gruen+Leisch:2008}.
If the mixture model specified is not identifiable this implies that
neither the parameter estimates used to characterise the clusters nor
the partitions derived can be uniquely determined thus rendering the
interpretation of results futile.  If the non-identifiability leads to
two or more isolated clearly differing solutions which are
observationally equivalent, the data does not allow to distinguish
between them. In this case domain knowledge might help in deciding
which of these solutions represents a useful clustering and allow to
exclude the alternative solutions.

Specific models might be useful if the multivariate data has a
restricted support, for instance, when the  observations or their squared values sum
to one,  in which case  the multivariate data points have the unit
simplex or the unit hypersphere  as support. Clustering data on the unit
hypersphere is sometimes also seen as clustering data only with
respect to the directions implied, while neglecting the length
information of the data points. Using a component distribution which
has as support the unit hypersphere leads for example to mixtures of
von Mises-Fisher distributions
\citep{model-based:Banerjee+Dhillon+Ghosh:2005}. This model has also
been shown to represent the corresponding probabilistic model for
spherical clustering where cosine similarity is used as similarity
measure in $k$-centroid clustering, similar to how Gaussian mixtures
are a probabilistic model for $k$-means clustering.

Analysing the topology of systems of interacting components is based
on network data where the components correspond to nodes and their
interactions to edges. This graph or network data represents a
specific data structure and is often stored using an adjacency
matrix. If the data contains $n$ nodes, the adjacency matrix is of
dimension $n \times n$ and the entries reflect if there are edges
between nodes together with the edge weights. In the simple case of a
symmetric, unweighted graph, the adjacency matrix is a symmetric
binary matrix. Clustering methods applied
to network data aim at finding community structures or at grouping
nodes together which are similar in their interactions
\citep{model-based:Handcock+Raftery+Tantrum:2007,
  model-based:Newman+Leicht:2007}.

\subsection{Combining components to clusters}\label{sec:comb-comp-corr}

While mixture  models as considered in Section~\ref{sec:comp-corr-clust}
often are able to capture the data distribution,   the assumption that there is a one-to-one relationship between
components and clusters might be violated.  Some components might be  too
\emph{similar} to form separate clusters and  should rather
be merged.

In this situation,  a mixture of mixtures model can be useful where
groups of components are combined to form a single cluster. The
inference on clusters can be either performed simultaneously with model
fitting or as a subsequent  step, after having fitted a
suitable model for approximating the data distribution. In the
following, several  strategies are discussed for multivariate continuous data and  some of these strategies might also be employed for other
types of multivariate data.

Finite mixtures of Gaussian distributions are not only a suitable
model for model-based clustering, but can also be applied for
semi-parametric approximation of general distribution functions. A
hierarchical mixture of mixtures of Gaussian distributions model can
thus be used in a situation where a single Gaussian distribution is
not flexible enough to capture the cluster distribution. In this
hierarchical model the components on the upper level of the mixture
correspond to clusters, while those on the lower level are only used
for semi-parametric estimation of the cluster distribution.

In its hierarchical structure the model is given by  
\begin{align*}
  p(y_i | \theta) & = \sum_{g=1}^G \eta_g f_g(y_i | \theta_g),\\
  f_g(y_i | \theta_g) & = \sum_{h=1}^{H_g} w_{gh} \phi(y_i | \mu_{gh}, \Sigma_{gh}).
\end{align*}
While such a model is appealing as it involves only Gaussian
distributions and thus is easy to understand and implement,
identifiability of the model is an issue because the likelihood is
invariant to assignment of components on the lower level to clusters
on the upper level.  In fact the density implied by this hierarchical
representation is equivalent to
\begin{align}\label{eq:mix-of-mix-flat}
  p(y_i | \theta) & = \sum_{g=1}^G \sum_{h=1}^{H_g}  \eta_g w_{gh}
  \phi(y_i | \mu_{gh}, \Sigma_{gh}).
\end{align}
This is to say that the hierarchical mixture of Gaussian mixtures
model is equivalent to a mixture of Gaussian distributions with
$\tilde{G}=\sum_{g=1}^G {H_g}$ components with weights
$\eta_{gh} = \eta_g w_{gh}$ and component-specific parameter vectors
$\mu_{gh}$ and $\Sigma_{gh}$.  Clearly the components in
Equation~\eqref{eq:mix-of-mix-flat} can be arbitrarily regrouped to
form clusters without changing the overall density.

To achieve identifiability in a hierarchical mixture of mixtures model
several approaches were proposed. They differ in that either
identification is already aimed for during model fitting or that first
a suitable semi-parametric approximation of the data distribution is
obtained using mixtures of Gaussian distributions and then a second
step is used for forming clusters by combining components.

\paragraph*{Direct inference for the mixture of mixtures approach.}

In order to directly fit this kind of model strong identifiability
constraints were imposed in a maximum likelihood framework.
\cite{model-based:Bartolucci:2005} considers only the case of
univariate data and specifies a mixture of Gaussian mixtures model
where the mixtures on the lower level are restricted to be unimodal and
to be the same for all clusters except for a mean shift.  A similar
restriction to have the same mixture distribution on the lower level
except for a mean shift was considered in
\citet{model-based:Dizio+Guarnera+Rocci:2007} for multivariate data.

Within a Bayesian framework the identifiability issue present due to
the invariance of the likelihood can be resolved by specifying
informative priors which allow to automatically distinguish between
components from the same and different clusters.
\citet{model-based:Malsiner-Walli+Fruehwirth-Schnatter+Gruen:2017}
consider this approach for finite mixtures, whereas
\citet{model-based:Yerebakan+Rajwa+Dundar:2014} use an infinite
mixture
approach. \citet{model-based:Malsiner-Walli+Fruehwirth-Schnatter+Gruen:2017}
proposed a prior structure which reflects the prior assumptions on the
separateness of the clusters and the compactness of their shapes and
which can be suitably adapted for different kinds of
applications. Thus the clustering notions, i.e., the cluster solutions
aimed for, are explicitly included in the mixture model using
informative priors in a Bayesian setting.  Their prior also implies
that the mixture on the lower level used for approximating the cluster
distributions is allowed to contain components with different
covariance matrices, but where the component means are pulled towards
a common mean corresponding to the centre of the cluster.  The prior
structure employed by \citet{model-based:Yerebakan+Rajwa+Dundar:2014}
is more rigid. In their approach the covariance matrices on the lower
level are assumed to be the same within a cluster and also that the mean
parameters on the lower level scatter according to a scaled version of
the same covariance matrix.

\paragraph*{Two-step procedures.}

In the first step mixtures of multivariate Gaussians are fitted as
semi-parametric approximations of the data distribution. The use of
multivariate Gaussian mixtures for the semi-parametric approximation
raises another issue. Often different Gaussian mixture models allow to
similarly well approximate the data distribution because the
approximation might either use only few components with complex
component distributions or many components with simple component
distributions. In case of Gaussian mixture models the complexity of the
component distribution is primarily governed by the structure of the
covariance matrix.

Subsequently a merging approach is employed in order to form
meaningful clusters given the Gaussian components. Different criteria
were proposed for deciding on merging in a stepwise procedure such as
closeness of the means \citep{model-based:Li:2005}, 
the modality of the resulting clusters
\citep{model-based:Chan+Feng+Ottinger:2008,
  model-based:Hennig:2010}, 
the entropy of the resulting partition
\citep{model-based:Baudry+Raftery+Celeux:2010}, 
the collocation of the observations
\citep{model-based:Molitor+Papathomas+Jerrett:2010}, the degree of
overlap measured by misclassification probabilities
\citep{model-based:Melnykov:2016}, and the the use of clustering cores
\citep{model-based:Scrucca:2016}.
 
This second step corresponds to another cluster analysis being
performed where the input are not the data points, but the estimated
mixture components.  The mixture is assumed to be too fine-grained to
represent a good cluster solution and groups of components need to be
formed to obtain the clusters.  Note that in particular those
approaches which only take the conditional probabilities of component
memberships or collocations of observations into account can directly
be used for any kind of mixture models to merge components to clusters
regardless of the distributions used for the components of the
mixture.

\subsection{Selecting the clustering base}\label{sec:select-clust-base}

The variables included as clustering base are in general assumed to
all equally contribute to the clustering solution. Each of the
variables is assumed to be in line with and reflect the cluster
structure. In general the variables used for cluster analysis should
be carefully selected because choosing a different set of variables
might change the meaning of the resulting cluster solution
\citep{model-based:Hennig:2015}.  Even efforts are made to select a
suitable clustering base based on theoretic considerations driven by
domain knowledge, the selected variables might not all be equally
useful to cluster the data or contribute equally to a selected
clustering solution.  In fact some of the variables included in the
cluster base might turn out to either contain no cluster structure or
be irrelevant for the obtained clustering solution. The inclusion of
irrelevant variables has several drawbacks. In particular their
inclusion complicates model selection due to overfitting and makes the
interpretation of the cluster solution a harder task than
necessary. In the worst case some variables might even perturb the
cluster structure detected. Such variables are referred to as
\emph{masking} variables and if included deteriorate the cluster
solution.

Alternatively the aim could be to reduce the clustering base in order
to eliminate redundant variables. This approach assumes that the
clustering information contained in a subset of the variables is
sufficient to characterise the cluster solution obtained using all
variables. This task thus requires to identify the minimal set of
variables necessary to identify the cluster solution. Using only the
smaller set then might get rid of collinearity problems, reduces the
number of variables which need to be collected for future analyses and
facilitates the interpretation by providing a core set of essential
variables.

A range of variable selection methods for clustering have been
proposed which can be used in the heuristic and / or the model-based
context and which can be either applied prior to, during or after the
cluster analysis to identify a suitable subset of variables
\citep{model-based:Gnanadesikan+Kettenring+Tsao:1995,
  model-based:Steinley+Brusco:2008}.

\paragraph{Prior filtering.}

Prior filtering investigates the distribution of single variables and
assesses how well suited they might be to reveal cluster structure in
the data. Variables with a very homogeneous distribution do clearly
not allow to extract clusters from the data. These variables can be
excluded from the clustering base before performing the cluster
analysis. Clearly for very high-dimensional data prior filtering is
appealing because this might allow to substantially reduce the
dimensionality of the clustering problem.

In particular for continuous variables, indices to assess the
\emph{clusterability} of a variable were developed. These
clusterability indices aim to determine whether a variable allows to
meaningfully cluster the observations or if the observations exhibit a
tendency to form into clusters based on a specific variable.
\citet{model-based:Steinley+Brusco:2008a}, for instance, propose to
use a variance-to-range ratio for initial screening of each variable
and to exclude variables where these ratios are small.

\paragraph{Variable selection.}
 
Deciding on a suitable variable set during finite mixture model
estimation often is complicated by the fact that the suitability of a
variable set depends on the number of clusters and the specific
component model used, e.g., the specification of the covariance
matrices in Gaussian mixture modelling. Thus the decision on the
variable set needs to be made simultaneously while deciding on a
suitable number of clusters and component model.

Heuristic methods for exploring the model space with respect to
different number of clusters and variable sets have been proposed in
\citet{model-based:Raftery+Dean:2006},
\citet{model-based:Maugis+Celeux+Martin:2009} and
\citet{model-based:Dean+Raftery:2010} for Gaussian mixture models and
latent class analysis using maximum likelihood estimation and  
the Bayesian information criterion (BIC) for model comparison.

Alternatively within a Bayesian framework
\citet{model-based:Tadesse+Sha+Vanucci:2005} propose to use reversible
jump Markov chain Monte Carlo (MCMC) methods in Gaussian mixture
modelling to move between mixture models with different numbers of
components while variable selection is accomplished by stochastic
search through the model space. In the context of infinite mixtures
\citet{model-based:Kim+Tadesse+Vannucci:2006} combine stochastic
search for cluster-relevant variables with a Dirichlet process prior
on the mixture weights to estimate the number of components.
\citet{model-based:White+Wyse+Murphy:2016} suggest to use collapsed
Gibbs sampling in the context of latent class analysis to perform
inference on the number of clusters as well as the usefulness of the
variables.

\paragraph{Shrinkage methods.}

A different approach to address the variable selection problem is to
use shrinkage or penalisation methods. In a frequentist setting these
methods correspond to maximising a penalised likelihood while in a
Bayesian setting suitable priors are used to induce shrinkage.  This
approach aims to induce solutions which favour a homogeneous
distribution over some variable or parameter in case the evidence for
heterogeneity is insufficient. This allows to prevent overestimating
heterogeneity and provides insights into which variables or parameters
are relevant for the cluster solution.

The shrinkage approaches pursued build on work in regression analysis
for variable selection. In this context for example the lasso (least
absolute shrinkage and selection operator;
\citealt{model-based:Tibshirani:1996}) was proposed to perform
variable selection. For the lasso penalty the penalised likelihood
estimate has exact zeros for some of the regression coefficients
instead of small values. Similar results are obtained for the maximum
a-posteriori estimate if the lasso is used as prior for the regression
coefficients. In the mixture context the lasso penalty or prior is
imposed on the parameter representing the difference between the
cluster value and the overall value of the parameter thus inducing
solutions where these differences are shrunken towards zero and the
overall value of the parameter is the same over all clusters.

In Gaussian mixture models shrinkage approaches have been proposed
which only impose the penalty or shrinkage prior on the cluster means.
This specification of the shrinkage reflects the assumption that the
component means characterise the clusters and thus are different for
variables relevant for clustering.  This approach was used in
\citet{model-based:Pan+Shen:2007} using penalised maximum likelihood
estimation. Using a fully Bayesian approach
\citet{model-based:Malsiner-Walli+Fruehwirth-Schnatter+Gruen:2016}
employed the normal-gamma prior for the differences in the component
means using finite mixtures and \citet{model-based:Yau+Holmes:2011}
imposed a double exponential prior on the differences in the component
means while fitting infinite mixtures.

\paragraph{Post-hoc selection.} This approach aims at arriving at a
cluster solution for a subset of variables which is similar to the
cluster solution obtained using all variables. This procedure is based
on the assumption that the clustering base does not contain any
masking variables, but the best cluster solution is obtained using all
variables. However, some of the variables might be redundant because
they either contain no or the same information on the cluster
structure than other variables in the segmentation base. In the
post-hoc selection step the aim is to identify these redundant
variables and omit them. After omitting these variables the cluster
problem has been recasted a a lower dimensional problem and results
are easier to interpret.
\citet{model-based:Fraiman+Justel+Svarc:2008} for example propose two
procedures to identify such variables assuming that a ``satisfactory''
grouping is given.

\subsection{Selecting the number of clusters}\label{sec:selecting-model}

Selecting the number of clusters is quite a controversial topic.
For finite mixtures, a suitable number of components can be selected
using different criteria. Information criteria such as the Akaike
information criterion (AIC) or Bayesian information criterion (BIC)
have been used in a model-based clustering context where it has been
shown for the BIC that the number of components are consistently
selected under certain conditions, in particular ensuring that the
component densities remain bounded \citep{model-based:Keribin:2000}.

If the model is fitted as part of the merging approach, there has been
less controversy around methods to select a suitable model. In this
case only a suitable semi-parametric approximation is required and the
determination of the number of clusters is made in the subsequent step.

In the case where components are assumed to correspond to clusters,
the situation is more complicated. Mixture models cannot only be used
to obtain a clustering, but also for semi-parametric density
approximation. Information criteria developed for general model
assessment usually aim at identifying a solution which reflects the
data-generating process well and thus rather select a model which
represents a well semi-parametric approximation of the data
distribution. The information criteria are ignorant of the clustering
purpose the model is finally used for.  In case where the component
distribution does not exactly match the cluster distribution, the
number of clusters tends to be overestimated using these
criteria. This problem becomes more severe the more data is
available. The larger the data set the better the cluster
distributions can be estimated and deviations from the imposed
parametric distribution are more severely penalised.

In order to improve the model selection performance when using mixture
models of parametric distributions for model-based clustering, an
alternative criterion was proposed. This criterion does not only
measure the suitability of a model based on the goodness-of-fit for
the data distribution, as indicated by the log-likelihood, but also
how well observations can be assigned to clusters. The later is
equivalent to having well separated cluster distributions where the
conditional probabilities of component membership unambiguously allow
to assign the observations to one of the components. Thus the entropy
of the conditional probabilities of component membership is also taken
into account leading to the integrated completed likelihood
information criterion
\citep[ICL;][]{model-based:Biernacki+Celeux+Govaert:2000}, i.e.,
\begin{align*}
  \text{ICL}(G) &= \sum_{i=1}^n \log f(y_i, \hat{z}_i | G, \hat{\theta}) - \frac{\upsilon_G}{2}\log n,
\end{align*}
where $\hat{z}_i$ is the estimated cluster membership for observation
$i$, $\hat{\theta}$ the estimated mixture parameters and $\upsilon_G$
the number of estimated parameters.  In contrast to AIC or BIC, ICL
aims at identifying well separated clusters. This avoids
overestimating the number of clusters by taking into account that the
estimated model is used to assign observations to clusters and to
obtain a suitable partition of the observations.

\section{Post-processing the fitted model}\label{sec:post-processing}

\subsection{Identifying the model}\label{sec:identifying-model}
Mixture models suffer from generic non-identifiability issues due to
label switching \citep{model-based:Redner+Walker:1984}, i.e., the
mixture distribution is the same regardless of which label is assigned
to which cluster. Only inference on label invariant quantities can be
easily accomplished while inference on quantities which depend on a
unique labelling require an identified mixture model.

If only point estimates of mixture models are used, such as  maximum
likelihood estimates in a frequentist framework or   maximum
a-posteriori estimates in a Bayesian setting, a unique solution is
easily obtained by imposing an ordering constraint. If uncertainty
estimates based on bootstrap techniques in frequentist estimation and
MCMC samples in Bayesian estimation with symmetric priors are to be
derived, the situation is more complicated and different methods for
obtaining an identified model have been proposed including
(a) imposing an ordering constraint \citep{fru:mcm},
(b) applying label-invariant loss functions in cluster and
  relabelling algorithms \citep{model-based:Stephens:2000},
(c) fixing the component membership of some observations
  \citep{model-based:Chung+Loken:Schafer:2004},
(d) relabelling with respect to the point estimate, e.g., the
  maximum a-posteriori estimate
  \citep{model-based:Marin+Mengersen+Robert:2005},
(e) clustering in the point process representation
  \citep{model-based:Fruehwirth-Schnatter:2006,fru:dea},
(f) probabilistic approaches taking the uncertainty of the
  relabelling into account \citep{model-based:Sperrin+Jaki+Wit:2010}.
For an overview see also
\citet{model-based:Jasra+Holmes+Stephens:2005} and
\citet{model-based:Papastamoulis:2016}.

\subsection{Determining a partition}\label{sec:determ-part} 

\cite{bin:bay} considered Bayesian cluster analysis as the task to
determine a suitable partition given a data set and assuming an
underlying mixture model. Thus more work has been done in the Bayesian
context to obtain a suitable partition based on $P(\bm{z} | \bm{y})$.
In the frequentist setting a partition is in general determined using
an identified model to classify observations separately and obtain a
partition.

In the Bayesian context \cite{bin:bay} proposed to obtain the optimal
partition by minimising the expected loss given $P(\bm{z} | \bm{y})$
and suggested several different loss functions. This approach
corresponds to minimising
\begin{align*}
  \mbox{E}(\ell(\hat{\bm{z}}, \bm{z}) | \bm{y})
\end{align*}
with respect to $\hat{\bm{z}}$ given the loss $\ell(\cdot, \cdot)$
between two classifications of the data.  One possible loss function
is to use the 0/1 loss, where
\begin{align*}
  \ell(\hat{\bm{z}}, \bm{z})
  &=
    \left\{
    \begin{array}{ll}
      0&\text{if }\hat{\bm{z}} = \bm{z},\\
      1&\text{if }\hat{\bm{z}} \neq \bm{z}.\\
    \end{array}\right.
\end{align*}
or a label-invariant version thereof. Using the 0/1 loss function
results in the maximum a-posteriori estimate, i.e., leads to selecting
the mode of $P(\bm{z} | \bm{y})$.  The drawback of the 0/1 loss is
that as long as two classifications are not the same they are assessed
as equally different by assigning a loss of one. This has lead to a
number of alternative loss functions being suggested.  Depending on
the loss function used in general either an identified model or the
posterior distribution of collocation is required to determine the
loss minimising classification.  For more details see also
\citet[Section~7.1.7]{model-based:Fruehwirth-Schnatter:2006}.

\paragraph{Based on an identified model.}

Given an identified model the cluster memberships can be inferred from
the latent allocation variables $z_i$ and conditional  on the
parameters $\theta$ of the mixture model. In this case the latent allocation
variables are independent and group memberships can be inferred
separately for each observation based on the conditional distributions
of cluster memberships $\zm=(z_1,\ldots,z_n)$  given the data $\ym=(y_1,\ldots,y_n)$,
\begin{align*}
  \Prob (z_1 = g_1, \ldots, z_n = g_n |\ym,   \theta) %
  \propto \prod_{i=1}^n \eta_{g_i} f_{g_i}(y_i |                              \theta_{g_i}),
\end{align*}
where $g_i$ is the cluster observation $y_i$ is assigned to.  To
determine a clustering different estimates can be used based on the
conditional probabilities of cluster memberships, e.g., observations
can be assigned to the cluster they are most likely from or assigned
by drawing from the conditional probabilities.

Assigning the observations to the cluster they are most likely from
results in the classification which is also obtained when minimising
the misclassification rate as loss function. This classification is
obtained by setting
\begin{align*}
  \hat{z}_i &= {\arg\,\max}_{g}\Prob(z_i = g | y_i, \theta)
\end{align*}
for each observation $i$.

\paragraph{Based on the collocation matrix.}

Partitions have the advantage that they are label-invariant
quantities. This implies that procedures for determining a final
partition can be used which do not require an identified model, but
only use the estimated collocation of observations as input. This also
implies that the only output required from model fitting is the
information on the partitions, i.e., collapsed sampling schemes can be
used in an MCMC context.

The collocation matrix is a $n\times n$ matrix of values between 0 and
1 with 1s in the diagonal where the $(i,j)$ entry represents the
probability that observations $i$ and $j$ are assigned to the same
component of the fitted mixture model, i.e., $P(z_i = z_j | \bm{y})$.
Several approaches for deriving a suitable partition from such a
collocation matrix have been proposed in a Bayesian setting where
draws from the posterior distribution of partitions are
available. These suggestions include (a) minimising a pairwise
coincidence loss function \citep{bin:bay,model-based:Lau+Green:2007},
(b) reformulating $P(z_i = z_j | \bm{y})$ as a dissimilarity matrix
and using partitioning around medoids
\citep[PAM;][]{model-based:Kaufman+Rousseeuw:1990}, a standard
partitioning clustering technique
\citep{model-based:Molitor+Papathomas+Jerrett:2010}, (c) determining a
partition which minimises the squared distance as loss function
\citep{model-based:Fritsch+Ickstadt:2009}, and (d) minimising the
variation of information as loss function
\citep{model-based:Wade+Ghahramani:2018}.  Of particular note is that
the partition minimising the posterior expected loss can in general
not be obtained directly, but iterative optimisation methods need to
be employed. Due to the size of the partition space this is a
computational demanding task.

\subsection{Characterising clusters}\label{sec:char-clust}  

Cluster analysis aims at determining groups in the data. The results
of a cluster analysis, however, in general not only consist of a
partition of the data, but also a characterisation of the groups. The
clustering together with the characterisation of the clusters allows
to summarise a multivariate data set and to give insights into its
structure. The characterisation of the clusters allows to profile them
and provides insights into the latent groups or types identified. This
can be used to describe the groups or even assign them meaningful
names.

In heuristic cluster analysis often a prototype for each group or
cluster is determined. In model-based clustering the cluster
distributions allow to characterise the clusters. If parametric
distributions are used for the clusters, often focus is given to
reporting and comparing the parameter estimates together with their
associated uncertainty.

Comparing the differences between prototypes in heuristic clustering
is not straightforward because the clustering algorithm aimed at
maximising these differences. Model-based methods thus allow to assess
differences between clusters using a sound statistical framework. This
facilitates to identify the variables which contribute most to the
clustering, and to assess if the resulting clustering is useful.  The
cluster distributions can either be determined based on an identified
mixture model or conditional on the partition or clustering obtained
for the data.

\subsection{Validating clusters}\label{sec:clust-valid}

General validation techniques have been developed for cluster analysis
which can be used regardless of the clustering method used. These
validation methods are distinguished into internal, external and
stability-based methods. Furthermore the assessment can be on the
level of the global cluster solution or specific to a single
cluster. A general overview on cluster validation methods is for
example given in
\citet{model-based:Halkidi+Batistakis+Vazirgiannis:2001}, while
\citet{model-based:Brock+Pihur+Datta:2008} provide an overview on
internal and stability-based measures as well as biological ones in
the context of bioinformatic applications.

\paragraph{Internal measures.}

The use of internal measures is appealing because they only require
data which is already available. However, they rely on a notion of
distance between observations. This notion is readily available when
heuristic clustering methods are applied. For the application of
model-based clustering methods no distance or dissimilarity measure
between observations needs to be specified. So this needs to be
additionally done in order to be able to calculate the internal
measures.

Most of the internal measures include information on within-cluster
scattering (i.e., compactness) and between-cluster
separation. Examples for these measures are silhouette width
\citep{model-based:Rousseeuw:1987}, the Dunn index
\citep{model-based:Dunn:1974} or the Davies-Bouldin index
\citep{model-based:Davies+Bouldin:1979}.  However, these measures are
in general only useful for convex shaped clusters and fail to provide
suitable insights into the quality of a cluster solution in cases of
arbitrarily shaped non-convex clusters and if noisy observations are
present. It is also obvious that if the internal criterion coincides
with the criterion minimised in the algorithm used for heuristic
clustering the solution obtained with this method should perform
well. This thus seems to be an unfair comparison. Nevertheless it
might still be useful to compare different cluster results on this
basis. This comparison provides insights into how much worse
alternative solutions are which are derived using a different cluster
criterion. Such an investigation might also not only provide insights
into the cluster solutions obtained, but also what kind of cluster
structure might naturally be contained in the data.

Also internal measures specifically developed for the mixture model
context have been proposed. \citet{model-based:Celeux+Soromenho:1996}
suggest to assess the suitability of a fitted mixture model to be used
for clustering based on the entropy of the conditional probabilities
of cluster memberships.  Because the final partition of the data is
derived from these conditional probabilities of cluster memberships, a
mixture model is more suitable for clustering if observations can be
unambiguously assigned to clusters. This measure also captures the
loss of information incurred by using only the estimated partition as
result and neglecting the uncertainty of cluster assignment.

\paragraph{External measures.}

External measures relate the partition derived to some external
structure, i.e., partition, imposed on the data. For instance, an
additional categorical variable is available which induces a partition
of the data, but has not been used in the cluster analysis. Such a
comparison assumes implicitly that the aim of the cluster analysis was
to arrive at a partition of the data which is close to this
partition. In general using cluster analysis to extract a partition
which corresponds to a partition induced by an observed categorical
variable is questionable. If the target variable is observed it would
seem more natural to use a classification or supervised learning
approach.

Standard methods for assessing classification accuracy, e.g., the
misclassification rate, can be employed to compare the class labels to
the cluster labels. However, this approach requires that a mapping
from cluster labels to class labels needs to be determined, which
might eventually not be straightforward in case where the number of
clusters and classes are different. One approach might be to choose
the mapping which maximises the classification accuracy criterion
employed. As an alternative, label-invariant measures can be employed
which determine the similarity between two partitions regardless of
any labels assigned to each of the groups contained in the
partitions. This is achieved by determining the numbers of pairs of
observations which are in the same group for both partitions, in
different groups for both partitions and in different groups in one
partition and in the same group for the other. Based on these numbers
of pairs the Rand index \citep{model-based:Rand:1971} or adjusted Rand
index \citep{model-based:Hubert+Arabie:1985}, the Jaccard coefficient
\citep{model-based:Jaccard:1912}, the Fowlkes and Mallows index
\citep{model-based:Fowlkes+Mallows:1983} among others can be derived
and used as validation measures. A further criterion used is purity
\citep{model-based:Zhong+Ghosh:2003} which assesses to which extent a
cluster only contains observations from the same class, i.e., this
criterion does not penalise splitting classes into several clusters,
but deteriorates if classes are merged into the same cluster.

\paragraph{Stability measures.}
External measures compare two partitions. These measures thus can also
be used to assess the stability of cluster solutions \citep[see for
example][]{model-based:Hennig:2007}. The extent to which a cluster
solution depends on a specific data set and how much it varies if a
new data set is used can be assessed based on bootstrapping. Pairs of
bootstrap samples are drawn from the data and clustered. These two
cluster solutions induce each a partition in the original data
set. The similarity between these partitions is determined using an
external measure and can be used to assess stability. Alternatively it
might also be of interest to assess stability of clustering solutions
if different cluster algorithms are employed.

\citet{model-based:Dolnicar+Leisch:2010} point out that these
stability assessments allow to infer if cluster solutions can at least
be constructed in a stable way in the case when natural clusters,
i.e., density clusters, are not contained in the data. They propose to
use stability as a criterion to select a suitable clustering solution
in case no natural clusters are contained in the data set.

\subsection{Visualising cluster solutions}\label{sec:visu-clust-solut}

Suitable visualisation methods allow to assess the cluster quality,
illustrate the cluster shapes and gain an insight into the cluster
distributions. Visualisation is of particular importance in the
clustering context as clustering is an exploratory data analysis tool.
Assessing the quality of a cluster solution based on visualisations is
also often necessary because of the difficulty to formally define the
clustering problem in a way which ensures that the obtained solutions
have the desired characteristics. Furthermore, input from domain
experts to validate and optimise a cluster solution might be easier to
obtain if they are able to assess a suggested solution using suitable
visualisations.

\paragraph{Assessing cluster quality.} Given different cluster
validation indices it might be easier to compare them and select a
suitable solution in dependence of their values using visualisation
methods. E.g., if a clear cluster structure is suspected in the data
an elbow or optimal value of the criterion might be visually
discernible and used for model selection.  In the merging components
to clusters context \citet{model-based:Baudry+Raftery+Celeux:2010}
suggest to plot the entropy values versus the number of clusters and
select the solution where there is a break point in a piecewise linear
fit.

An additional visualisation technique based on an internal cluster
validation index is the silhouette plot
\citep{model-based:Rousseeuw:1987}. The silhouette plot illustrates
the quality of the cluster solution based on the silhouette values
grouped by cluster. As an alternative \citet{model-based:Leisch:2010}
proposed the shadow plot which has the same structure but uses the
shadow value instead of the silhouette value. The shadow value has the
advantage that it is computationally less demanding to determine than
the silhouette value. The drawback is that this index relies more on
the cluster centroid being a good representative.

\paragraph{Illustrating cluster shapes and separation.} Cluster shapes
and separateness can be illustrated using scatter plots of the data
points, at least for continuous data. However, in the case of
high-dimensional data this might not be a feasible approach and lower
dimensional representations of the data might be more useful. The
lower dimensional representations could either be based on general
techniques for dimension reduction such as principle component
analysis (PCA) or specific techniques for cluster analysis.  In the
context of Gaussian mixture modelling \citet{model-based:Scrucca:2010}
proposes to determine the subspace which captures most of the
cluster structure contained in the data. Alternatively also
cluster-specific projections were proposed which for a given cluster
maximises its distance to the other clusters
\citep{model-based:Hennig:2004}.

\paragraph{Characterising cluster prototypes.} Profile plots of the
prototypes can help to quickly identify in which variables clusters
differ and how they can be characterised
\citep{model-based:Dolnicar+Leisch:2014}.  Profile plots use the
information on the prototypes and visualise them. In the model-based
context this information consist of characterisations of the cluster
distributions, e.g., in case of parametric distributions their
parameters. Profile plots are based on conditional plots and
separately visualise each of the cluster results, but allow for easy
comparison. These plots can also be enhanced to include uncertainty
information. 

In the context of shrinkage priors imposed on the cluster means
\citet{model-based:Yau+Holmes:2011} and
\citet{model-based:Malsiner-Walli+Fruehwirth-Schnatter+Gruen:2016}
propose to visualise the posterior distribution of the shrinkage
factors for each of the variables using boxplots. Such a visualisation
indicates the variables for which the clusters differ and thus can be
used to characterise the clusters.

\paragraph{Visualising the example data sets.}

The three data sets introduced in Section~\ref{sec:spec-clust-probl}
and shown in Figures~\ref{fig:clustering-concepts} and
\ref{fig:clustering-concepts-with-cl} are used to create silhouette
plots and to visualise in a silhouette-type plot the cluster
uncertainty inherent in fitted suitable mixture models.

The silhouette value for an observation $i$ is given by
\begin{align*}
  s(i) &= \frac{b(i) - a(i)}{\max\{a(i), b(i)\}},
\end{align*}
where
\begin{align*}
  a(i) &= \frac{1}{n_{z_i}}\sum_{j \in \partset_{z_i}}\|y_i - y_j\|,\\
  b(i) &= \min_{g \in \{1, \ldots, G\} \setminus z_i} \frac{1}{n_g} \sum_{j \in \partset_g} \|y_i - y_j\|,
\end{align*}
i.e., $a(i)$ is the average Euclidean distance of observation $i$ to
all observations $j$ assigned to the same cluster and $b(i)$ is the
minimum average Euclidean distance of observation $i$ to observations
$j$ which are assigned to a different cluster. This definition ensures
that $s(i)$ takes values in $[-1, 1]$, where values close to one
indicate ``better'' clustering solutions.

\citet{model-based:Rousseeuw:1987} suggests to visualise the
silhouette values by grouping them by cluster and sorting them in
decreasing order within cluster. The width of the values for each
cluster indicates the size of the cluster and the distribution of
silhouette values within cluster indicates how well separated that
cluster is from the other clusters in Euclidean space. The average
silhouette value within a cluster serves as an indicator how compact
and well separated this cluster is. The overall cluster solution can
be assessed using the total average of silhouette values.

\begin{figure}[t!]
  \centering
  \includegraphics[width=0.7\textwidth]{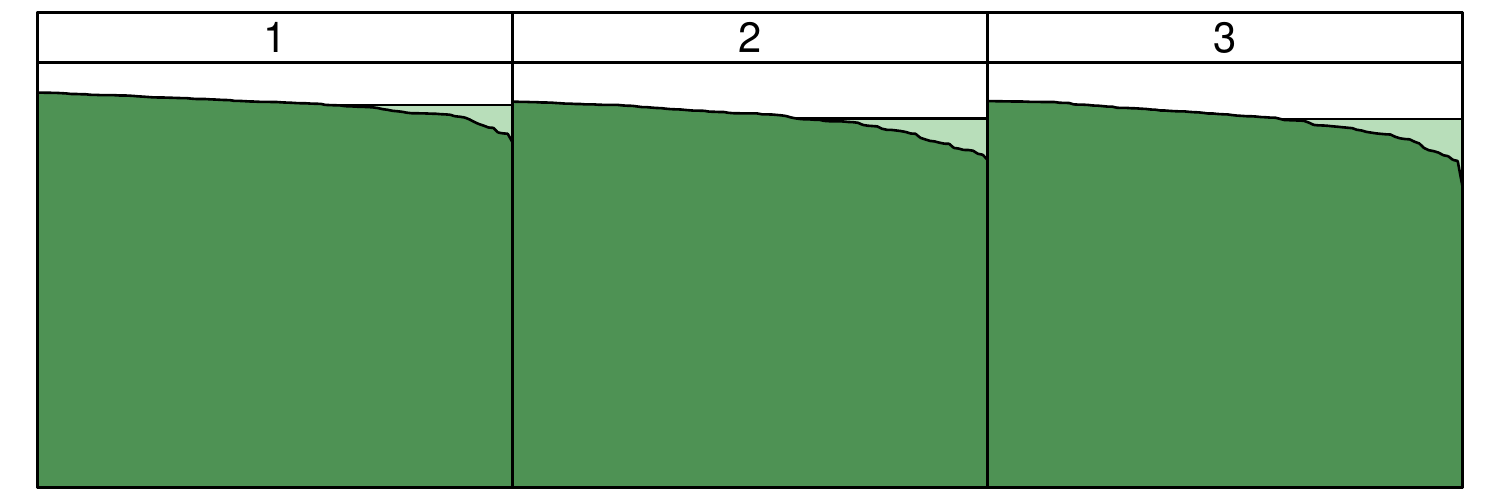}\\
  \includegraphics[width=0.7\textwidth]{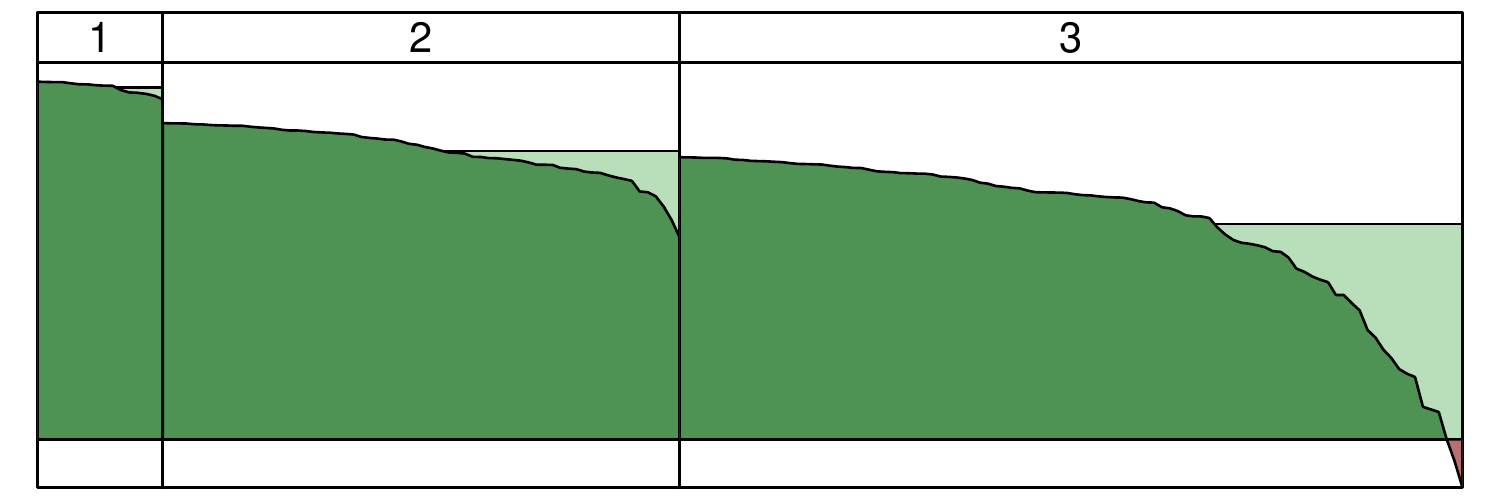}\\   
  \includegraphics[width=0.7\textwidth]{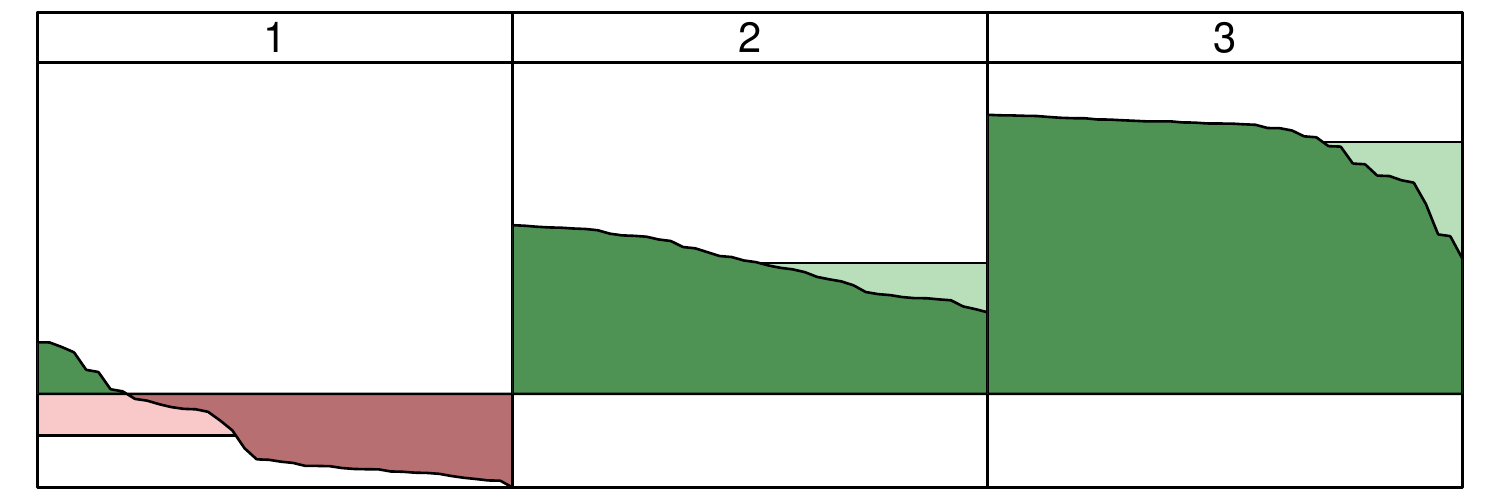}
  \caption{Silhouette plots based on the Euclidean distance and the
    true clustering for the three example data sets shown in
    Figure~\ref{fig:clustering-concepts}: (a) compact clusters (on the
    top); (b) density-based clusters (in the middle); (c) connected
    clusters or clusters sharing a functional relationship (on the
    bottom).}
  \label{fig:silhouettes}
\end{figure}
\begin{figure}[t!]
  \includegraphics[width=\textwidth]{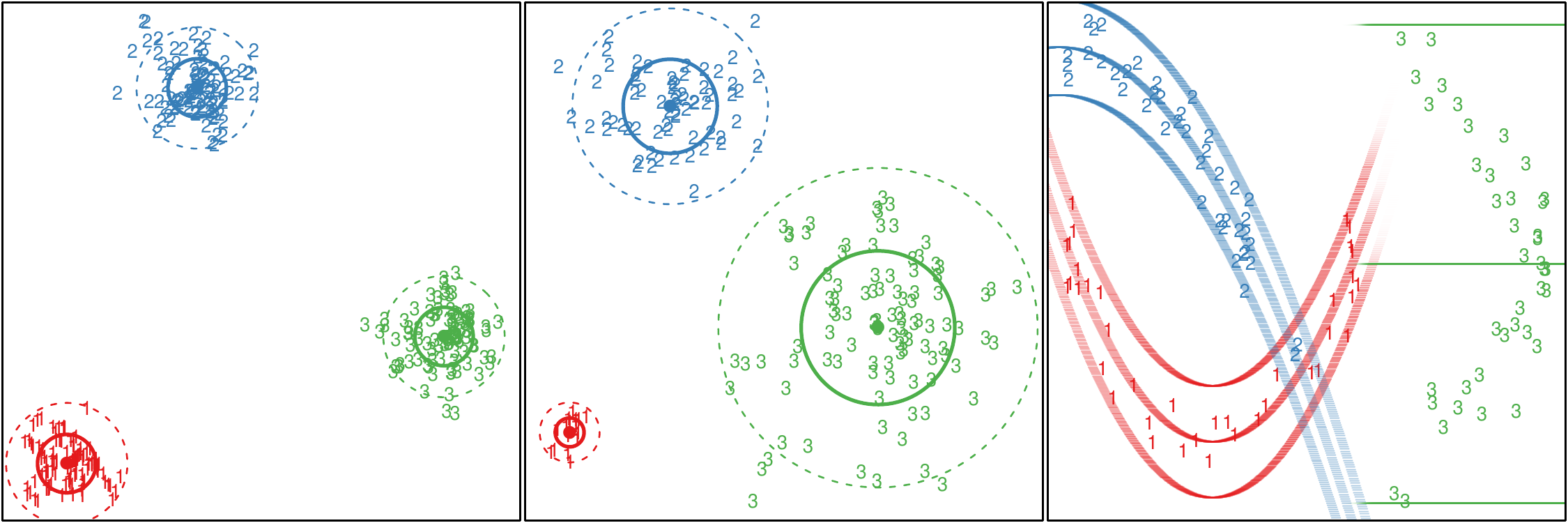}
  \caption{Plots of the example data sets together with the fitted
  mixture models: (a) a mixture of three Gaussian distributions with identical spherical
  covariance matrices (on the left);
  (b) a mixture of three Gaussian distributions with different spherical covariance matrices (in the middle);
  (c) a mixture of linear regression models with concomitant variables (on the right).}
  \label{fig:mixtures}
\end{figure}

Figure~\ref{fig:silhouettes} shows the silhouette values using
Euclidean distance and the true clustering solution for the three
artificial data sets. The plot on the top gives the silhouette plot in
the case compact clusters are present. The result indicates that
these clusters are well separated in Euclidean space and the plot also
reflects that the clusters are equally sized. For the case of
density-based clusters the silhouette plot indicates that the clusters
are of different size and the different levels of compactness of the
clusters impact the silhouette values within clusters (middle
plot). For the case of connected clusters which might be modelled using
some functional relationship the silhouette plot at the bottom
indicates that centroid-based partitioning methods using the Euclidean
distance might not be able to detect the true solution because most
observations in the first cluster, i.e., the $U$-shaped cluster in
Figures~\ref{fig:clustering-concepts} and
\ref{fig:clustering-concepts-with-cl} on the right, are in Euclidean
space closer on average to observations from a different cluster than
their own.

If model-based clustering methods are used to fit the different data
sets one can use: for data set (a) a mixture of Gaussian distributions
with identical spherical covariance matrices, for data set (b) a
mixture of Gaussian distributions with spherical covariance matrices
differing in volume and for data set (c) a mixture of linear
regression models with a horizontal line for the first cluster and
polynomial regressions of degree two for the other two clusters in
combination with a concomitant variable model (see also
Section~\ref{sec:mark-determ-mark}) based on the variable on the
$x$-axis, i.e., the cluster sizes dependent on variable $x$ in the
form of a multinomial logit model. The models obtained when fitted
using the EM algorithm initialised in the true solution are shown in
Figure~\ref{fig:mixtures}. For the first two examples the cluster
means are indicated together with the 50\% and 95\% prediction
ellipsoids (neglecting the uncertainty with which the parameters are
estimated) for the fitted components given by circles using full and
dashed lines respectively. For the third example the fitted regression
lines for each of the components are shown together with 95\%
pointwise prediction bands (neglecting the uncertainty with which the
parameters are estimated) and using an alpha-shading (i.e., a
transparency value) corresponding to the component size derived from
the concomitant variable model. The points are numbered  according to
their component assignments based on the maximum conditional
probabilities of cluster memberships obtained from the fitted mixture
models.

\begin{figure}[t!]
  \centering
  \includegraphics[width=0.7\textwidth]{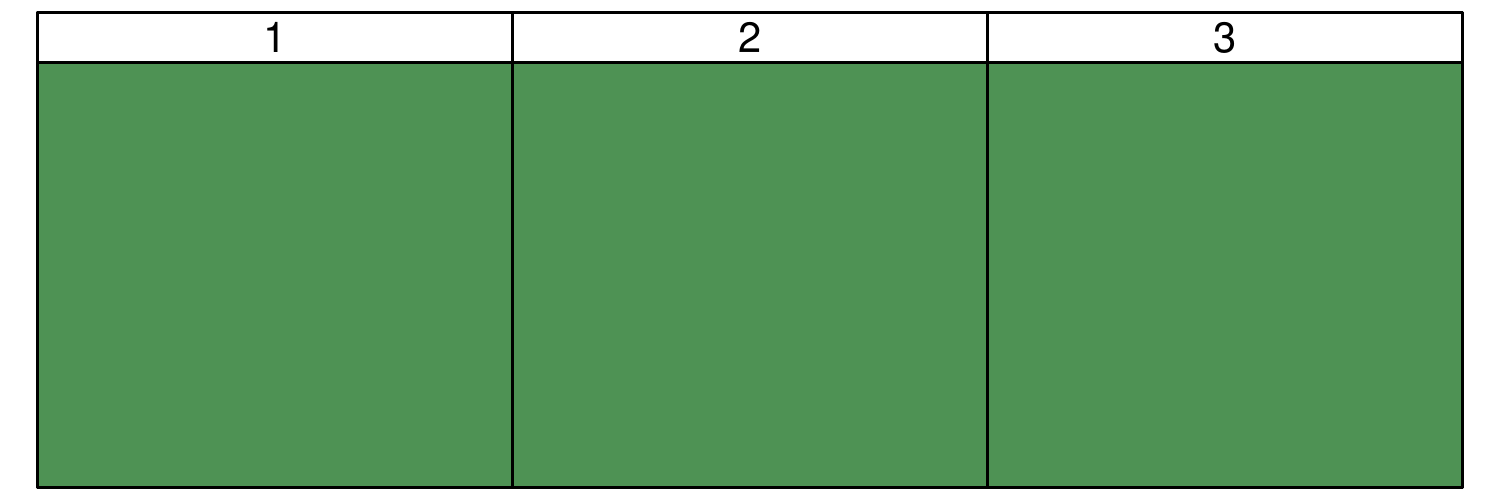}\\
  \includegraphics[width=0.7\textwidth]{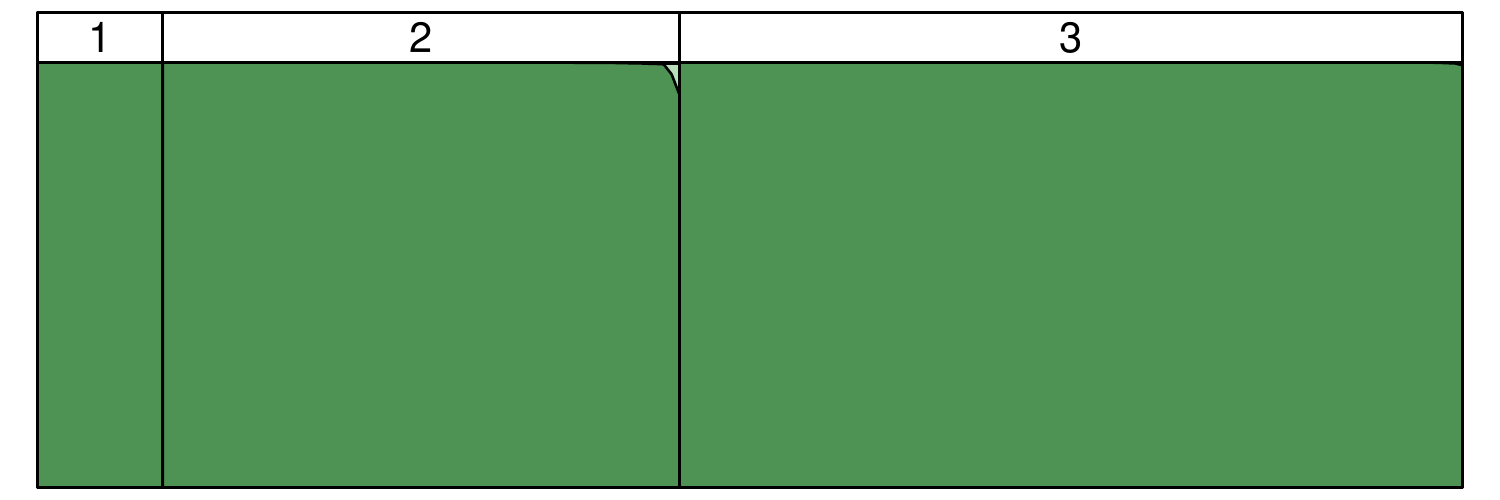}\\
  \includegraphics[width=0.7\textwidth]{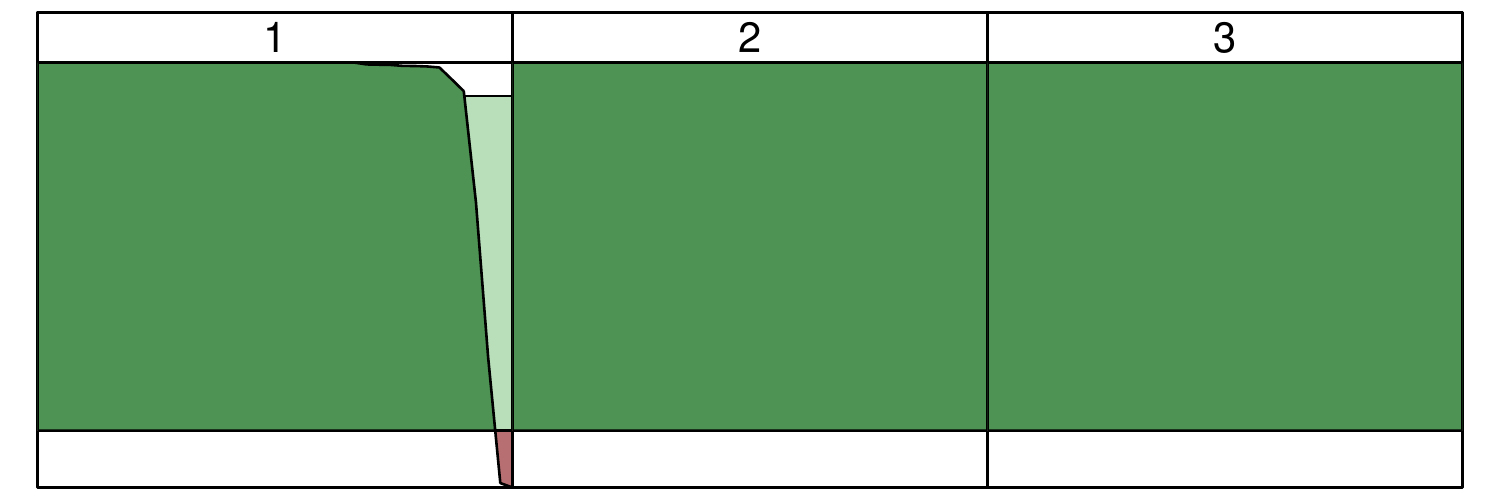}
  \caption{Plots of the conditional probabilities of cluster
    membership based on a finite mixture model and split by the true
    clustering for the three example data sets shown in
    Figure~\ref{fig:clustering-concepts}: (a) compact clusters (on the
    top); (b) density-based clusters (in the middle); (c) connected
    clusters or clusters sharing a functional relationship (on the
    bottom).}
  \label{fig:posteriors}
\end{figure}   

The conditional probabilities of cluster memberships obtained for the
three fitted mixture models split by the true cluster memberships are
visualised in Figure~\ref{fig:posteriors}. In this case clearly the
fitted mixture models -- even though not representing the true cluster
generating mechanism -- allow to identify the cluster memberships very
well.

\section{Illustrative applications}\label{sec:applications}

The areas of application are diverse and specialised model-based
clustering methods have been developed to meet needs and challenges
encountered in the different fields. The challenges were encountered
due to specific data structures or specific desired cluster solution
characteristics. Specific data structures are for example very
high-dimensional data or the availability of time series panel data.
Specific cluster solution characteristics are required if specific
cluster shapes or the presence of very small clusters are suspected in
the data.

\subsection{Bioinformatics: Analysing gene expression data}

In bioinformatics cluster methods have in particular emerged as useful
tools for analysing gene expression data; for an introduction see
\citet{model-based:McLachlan+Do+Ambroise:2004}. The aims in this area
are to reduce data dimensionality because of the large number of genes
present in the data, to verify if gene expression patterns differ
between observed groups using an unsupervised learning approach (i.e.,
a clustering instead of a classification approach where the group
labels are included in the analysis;
\citealt{model-based:Kebschull+Demmer+Gruen:2014}) to avoid
overfitting, and detect latent groups potentially being present.

In contrast to other applications analysing gene expression data poses
specific problems where model-based clustering was adapted in suitable
ways to provide better results than standard clustering methods. First
of all the data structure is different because in general the number
of observations is small compared to the number of dimensions. Thus
parsimonious Gaussian mixture models
\citep{model-based:McNicholas+Murphy:2008} based on mixtures of factor
analysers \citep{model-based:McLachlan+Peel:2000,
  model-based:McLachlan+Peel+Bean:2003,
  model-based:McNicholas+Murphy:2010} emerged as a useful model-based
clustering method in this context. Furthermore time-course data led to
the extension of mixtures of linear models to mixtures of linear mixed
models using semi-parametric regression methods
\citep{model-based:Luan+Li:2003, model-based:Luan+Li:2004,
  model-based:Celeux+Martin+Lavergne:2005,
  model-based:Ng+McLachlan+Wang:2006,
  model-based:Scharl+Gruen+Leisch:2010,
  model-based:Gruen+Scharl+Leisch:2012}.

In general also the distribution of the latent groups is in general
suspected to be neither isotropic nor symmetric and to contain extreme
observations and thus a single Gaussian distribution is not able to
capture the cluster distribution. In order to allow for more flexible
shapes mixtures of $t$ distributions were considered to allow for
heavier tails and skew distributions to account for non-symmetry
\citep{model-based:Pyne+Hu+Wang:2009,
  model-based:Fruehwirth-Schnatter+Pyne:2010,
  model-based:Franczak+Browne+McNicholas:2014,
  model-based:Vrbik+McNicholas:2012, model-based:Lee+McLachlan:2014,
  model-based:Hagan+Murphy+Gormley:2016}.

\subsection{Marketing: Determining market segments}\label{sec:mark-determ-mark}

In market research clustering methods are used for market
segmentation; for an introduction see for example
\citet{model-based:Wedel+Kamakura:2001}. Market segmentation is a key
instrument in strategic marketing. Market segmentation aims at
dividing the consumer or business market into sub-groups. These
sub-groups can then be targeted separately which provides competitive
advantages. Market segments need to fulfil certain criteria in order
to be useful in practice: identifiability, substantiality,
accessibility, stability, responsiveness, actionability. Some of these
criteria might even be seen as knock-out criteria such that clustering
solutions which do not comply with them cannot be considered for
implementing a successful market segmentation strategy. 

As pointed out by \citet{model-based:Allenby+Rossi:1999} there is no
clear consensus how consumer heterogeneity is modelled best. While
there is agreement that consumers differ in their interests,
preferences, etc., it is less clear if these heterogeneities are due
to the presence of latent groups or because of continuous individual
differences.  While latent groups would indicate the use of mixture
models, continuous differences might be better captured by a random
effects model. However, even if a random effects model is assumed to
be better suited, it is still doubtful that consumer heterogeneity
might be captured by a single Gaussian distribution. In general the
random effects distribution is not known and the assumption of a
Gaussian distribution, i.e., a symmetric unimodal distribution, will
be questionable. In this case the random effects distribution could
also either be approximated by a mixture distribution \citep[see for
example][]{model-based:Aitkin:1999} or the combination of a finite
mixture with random effects within the components. The later model is
also referred to as heterogeneity model
\citep{model-based:Fruehwirth-Schnatter+Tuechler+Otter:2004,
  model-based:Fruehwirth-Schnatter+Tuechler+Otter:2004a}.

\citet{model-based:Dolnicar+Leisch:2010} also argue that density
clusters rarely exist in consumer data and that the latent group model
assuming homogeneity within the groups is hardly ever well
fitting. They nevertheless defend the use of market segmentation and
the extraction of sub-groups. From a managerial point of view grouping
consumers into segments can still be beneficial and useful for targeting
and positioning, even if these groups do not reflect natural clusters
present in the data.

Specific extensions of mixture models developed for market
segmentation are mixtures of regression models
\citep{model-based:Wedel+DeSarbo:1995}. These models were useful to
extract market segments where the members have similar price
sensitivity values or respond in a similar way to promotions.
Previous to mixtures of regression models a heuristic method referred
to as clusterwise linear regression \citep{model-based:Spaeth:1979}
was used to obtain clusters where members are similar in the
regression coefficients. In clusterwise linear regression an
algorithmic approach to partition observations is pursued where linear
regression models are fitted within each group to minimise a
sum-of-squares criterion. Mixtures of regressions embed this approach
in a probabilistic framework.  Compared to clusterwise linear
regression mixtures of regressions have the advantages that they allow
to easily extend clustering of linear regressions to the clustering of
generalised linear models and even more general regression models.

Market segments are often defined using behavioural variables as
segmentation base because the intention is to form groups of consumers
who are similar in their behaviour. However, to ensure accessibility
these identified market segments also need to differ with respect to
other variables, e.g., socio-demographic information. To improve the
accessibility of the obtained market segmentation solution concomitant
variable models \citep{model-based:Dayton+Macready:1988} have been
employed \citep{model-based:Wedel+DeSarbo:2002}. While the cluster
structure is still determined based on the behavioural variables only,
the cluster memberships vary with the concomitant variables. This
means that for example multinomial logit models are used to model the
cluster memberships in dependence of the socio-demographic
variables. This allows to identify for which of the concomitant
variables the cluster memberships are significantly different and to
profile the segments.

\subsection{Psychology and sociology: Revealing latent structures}

Latent structure analysis evolved in psychology to model dependency
structures between observed variables; for a survey see
\citet{model-based:Andersen:1982}. Latent class analysis assumes that
the dependencies are caused through a latent variable which groups the
observations, i.e., conditional on the latent group the variables are
independent. For an introduction to latent class analysis see
\citet{model-based:Collins+Lanza:2010}.

In the case of binary variables this implies that within each group
the observations are from a product of Bernoulli distributions and the
differences in success probabilities between the groups lead to
dependencies observed in the aggregate data
\citep{model-based:Goodman:1974}. In psychology these latent groups
are associated with different types of respondents leading to
different prototypical answers on dichotomous item batteries.
Extension to ordinal variables also exist
\citep{model-based:Linzer+Lewis:2011}. In ordinal latent class
analysis the conditional independence assumption is retained and for
each dimension a different distribution for the ordinal variable is
assumed for each group.  An issue when applying this model class is
identifiability. In general these latent class models are not
generically identifiable because only local identifiability can be
ensured, but not global identifiability. Local identifiability implies
that in a neighbourhood of a parameterisation no other
parameterisation exists implying the same mixture distribution. The
lack of global identifiability in this case implies that there might
exist a different parameterisation of the same distribution which is
isolated from the other solution somewhere else in the parameter
space.

Areas of application arise where behavioural variables, attitudes or
values are collected on a dichotomous or categorical scale and the
observed values are associated because of latent traits in the
population which lead to jointly observing high or low values for a
group of variables. Latent class analysis has for example been used
for survey data on health and risk behaviour among youth to identify
and characterise different risk groups in the population
\citep[see][]{model-based:Collins+Lanza:2010}.

\subsection{Economics and finance: Clustering times series}

In economics the model-based clustering approach has proven useful by
allowing to account for unobserved heterogeneity in standard
econometric models which if neglected might lead to biased results and
thus to wrong conclusions being drawn. For example,
\citet{model-based:Alfo+Trovato+Waldmann:2008} investigate if growth
can be considered exogenous in the Solovian sense. They allow for
heterogeneity between countries in order to obtain a grouping of
countries based on their estimated model.

Specific applications of interest emerged for the analysis of time
series data; for an overview see for example
\citet{model-based:Fruehwirth-Schnatter:2011}. For instance, mixtures
of Markov switching models were developed to allow groups of time
series to follow a different Markov switching model
\citep{fru-kau:mod}. For an application to financial data consisting
of time series on returns from 21 European stock markets see for
example \citet{model-based:Dias+Vermunt+Ramos:2015} who identify three
clusters in the data where the Markov switching differs. Mixtures of
Markov chain models for categorical time series models are considered
in \citet{model-based:Fruewirth-Schnatter+Pamminger+Weber:2012} and
\citet{model-based:Fruewirth-Schnatter+Pamminger+Weber:2016}.
\citet{model-based:Fruewirth-Schnatter+Pamminger+Weber:2012} analyse
earnings development of young labour market entrants and
\citet{model-based:Fruewirth-Schnatter+Pamminger+Weber:2016}
investigate career paths of Austrian women after their first birth.

\subsection{Medicine and biostatistics: Unobserved heterogeneity}

Specific models developed and primarily used in medicine and
biostatistics are those concerned with survival analysis. Model-based
clustering has been employed in this area by extending the standard
models for survival analysis to the mixture case such as mixtures of
proportional hazard models
\citep{model-based:Rosen+Tanner:1999}.

Functional data is also often encountered in medical and
biostatistical applications. Functional linear models relate a
functional predictor with a scalar response by determining a smooth
regression parameter function where the integral of the functional
predictor times the parameter function over time equals the
conditional mean of the scalar response. The generalisation of this
functional regression model to mixtures is investigated in
\citet{model-based:Yao+Fu+Lee:2011}. Its application is illustrated on
egg-laying data from a fertility study where the functional regression
is used to relate fertility of the early life period to the total
lifetime and to gain insights into two distinct mechanisms relating
longevity and early fertility.

In the case of longitudinal data growth mixture modelling has been
considered \citep{model-based:Muthen+Asparouhov:2015,
  model-based:Muthen+Brown+Masyn:2002}. Growth mixture modelling refers
to the use of mixtures of generalised linear mixed models to
longitudinal data which allow to capture the intra-individual
dependencies over time as well as differences between individuals.

A further model-based clustering application in health research uses
Dirichlet process mixtures to assess performance of health centres and
classify them \citep{model-based:Ohlssen+Sharples+Spiegelhalter:2006,
  model-based:Zhao+Shi+Shearon:2015}. The Dirichlet process clustering
approach is used as a semi-parametric approach to approximate the
latent heterogeneity distribution over health centres. The clustering
structure obtained allows to identify groups of particularly badly or
well performing centres.

\section{Conclusion}\label{sec:summary}

Model-based clustering has emerged as a useful tool to perform cluster
analysis. The flexibility of this approach stems from the fact that
any statistical distribution or model can be used for the components.
This allows to come up with clustering techniques for any kind of data
where a statistical model is available for. However, this flexibility
also has drawbacks. The cluster structure aimed for might not
necessarily be reflected in a model-based clustering solution. This is
due to the difficulty to explicitly include the notions of the
clustering aimed for in the specified model. While the specified model
may capture the targeted cluster structure, also solutions are
possible which are not useful in a clustering context.  The selection
of a suitable clustering method and solution thus remains ambiguous
and might be perceived as lacking scientific rigour. Future
developments in model-based clustering thus hopefully address these
issues and ameliorate these problems. Furthermore, new areas of
application might arise which induce new methodological developments
and push the boundaries of this modelling technique further.

\bibliography{bg}

\begin{thebibliography}{135}
\newcommand{\enquote}[1]{``#1''}
\providecommand{\natexlab}[1]{#1}
\providecommand{\url}[1]{\texttt{#1}}
\providecommand{\urlprefix}{URL }
\expandafter\ifx\csname urlstyle\endcsname\relax
  \providecommand{\doi}[1]{doi:\discretionary{}{}{}#1}\else
  \providecommand{\doi}{doi:\discretionary{}{}{}\begingroup
  \urlstyle{rm}\Url}\fi
\providecommand{\eprint}[2][]{\url{#2}}

\bibitem[{Aitkin(1999)}]{model-based:Aitkin:1999}
Aitkin M (1999).
\newblock \enquote{A General Maximum Likelihood Analysis of Variance Components
  in Generalized Linear Models.}
\newblock \emph{Biometrics}, \textbf{55}(1), 117--128.
\newblock \doi{10.1111/j.0006-341x.1999.00117.x}.

\bibitem[{Alf{\'o} \emph{et~al.}(2008)Alf{\'o}, Trovato, and
  Waldmann}]{model-based:Alfo+Trovato+Waldmann:2008}
Alf{\'o} M, Trovato G, Waldmann RJ (2008).
\newblock \enquote{Testing for Country Heterogeneity in Growth Models Using a
  Finite Mixture Approach.}
\newblock \emph{Journal of Applied Econometrics}, \textbf{23}(4), 487--514.
\newblock \doi{10.1002/jae.1008}.

\bibitem[{Allenby and Rossi(1999)}]{model-based:Allenby+Rossi:1999}
Allenby GM, Rossi PE (1999).
\newblock \enquote{Marketing Models of Consumer Heterogeneity.}
\newblock \emph{Journal of Econometrics}, \textbf{89}(1--2), 57--78.
\newblock \doi{10.1016/s0304-4076(98)00055-4}.

\bibitem[{Andersen(1982)}]{model-based:Andersen:1982}
Andersen EB (1982).
\newblock \enquote{Latent Structure Analysis: A Survey.}
\newblock \emph{Scandinavian Journal of Statistics}, \textbf{9}(1), 1--12.

\bibitem[{Azzalini and {Dalla
  Valle}(1996)}]{model-based:Azzalini+Dalla-Valle:1996}
Azzalini A, {Dalla Valle} A (1996).
\newblock \enquote{The Multivariate Skew-Normal Distribution.}
\newblock \emph{Biometrika}, \textbf{83}(4), 715--726.
\newblock \doi{10.1093/biomet/83.4.715}.

\bibitem[{Azzalini and Menardi(2014)}]{model-based:Azzalini+Menardi:2014}
Azzalini A, Menardi G (2014).
\newblock \enquote{Clustering via Nonparametric Density Estimation: The {R}
  Package {pdfCluster}.}
\newblock \emph{Journal of Statistical Software}, \textbf{57}(11), 1--26.
\newblock \doi{10.18637/jss.v057.i11}.

\bibitem[{Banerjee \emph{et~al.}(2005)Banerjee, Dhillon, Ghosh, and
  Sra}]{model-based:Banerjee+Dhillon+Ghosh:2005}
Banerjee A, Dhillon IS, Ghosh J, Sra S (2005).
\newblock \enquote{Clustering on the Unit Hypersphere Using Von
  {M}ises-{F}isher Distributions.}
\newblock \emph{Journal of Machine Learning Research}, \textbf{6}(September),
  1345--1382.

\bibitem[{Banfield and Raftery(1993)}]{model-based:Banfield+Raftery:1993}
Banfield JD, Raftery AE (1993).
\newblock \enquote{Model-Based {G}aussian and Non-{G}aussian Clustering.}
\newblock \emph{Biometrics}, \textbf{49}(3), 803--821.
\newblock \doi{10.2307/2532201}.

\bibitem[{Bartolucci(2005)}]{model-based:Bartolucci:2005}
Bartolucci F (2005).
\newblock \enquote{Clustering Univariate Observations via Mixtures of Unimodal
  Normal Mixtures.}
\newblock \emph{Journal of Classification}, \textbf{22}(2), 203--219.
\newblock \doi{10.1007/s00357-005-0014-7}.

\bibitem[{Baudry \emph{et~al.}(2010)Baudry, Raftery, Celeux, Lo, and
  Gottardo}]{model-based:Baudry+Raftery+Celeux:2010}
Baudry JP, Raftery A, Celeux G, Lo K, Gottardo R (2010).
\newblock \enquote{Combing Mixture Components for Clustering.}
\newblock \emph{Journal of Computational and Graphical Statistics},
  \textbf{2}(19), 332--353.
\newblock \doi{10.1198/jcgs.2010.08111}.

\bibitem[{Biernacki \emph{et~al.}(2000)Biernacki, Celeux, and
  Govaert}]{model-based:Biernacki+Celeux+Govaert:2000}
Biernacki C, Celeux G, Govaert G (2000).
\newblock \enquote{Assessing a Mixture Model for Clustering with the Integrated
  Completed Likelihood.}
\newblock \emph{IEEE Transactions on Pattern Analysis and Machine
  Intelligence}, \textbf{22}(7), 719--725.
\newblock \doi{10.1109/34.865189}.

\bibitem[{Biernacki \emph{et~al.}(2003)Biernacki, Celeux, and
  Govaert}]{model-based:Biernacki+Celeux+Govaert:2003}
Biernacki C, Celeux G, Govaert G (2003).
\newblock \enquote{Choosing Starting Values for the {EM} Algorithm for Getting
  the Highest Likelihood in Multivariate {G}aussian Mixture Models.}
\newblock \emph{Computational Statistics \& Data Analysis}, \textbf{41}(3--4),
  561--575.
\newblock \doi{10.1016/s0167-9473(02)00163-9}.

\bibitem[{Binder(1978)}]{bin:bay}
Binder DA (1978).
\newblock \enquote{Bayesian Cluster Analysis.}
\newblock \emph{Biometrika}, \textbf{65}(1), 31--38.
\newblock \doi{10.1093/biomet/65.1.31}.

\bibitem[{Brock \emph{et~al.}(2008)Brock, Pihur, Datta, and
  Datta}]{model-based:Brock+Pihur+Datta:2008}
Brock G, Pihur V, Datta S, Datta S (2008).
\newblock \enquote{{clValid}: An {R} Package for Cluster Validation.}
\newblock \emph{Journal of Statistical Software}, \textbf{25}(4), 1--22.
\newblock \doi{10.18637/jss.v025.i04}.

\bibitem[{Browne and McNicholas(2012)}]{model-based:Browne+McNicholas:2012}
Browne RP, McNicholas PD (2012).
\newblock \enquote{Model-Based Clustering, Classification, and Discriminant
  Analysis of Data with Mixed Type.}
\newblock \emph{Journal of Statistical Planning and Inference},
  \textbf{142}(11), 2976--2984.
\newblock \doi{10.1016/j.jspi.2012.05.001}.

\bibitem[{Bryant and Williamson(1978)}]{model-based:Bryant+Williamson:1978}
Bryant P, Williamson JA (1978).
\newblock \enquote{Asymptotic Behaviour of Classification Maximum Likelihood
  Estimates.}
\newblock \emph{Biometrika}, \textbf{65}(2), 273--281.
\newblock \doi{10.1093/biomet/65.2.273}.

\bibitem[{Cai \emph{et~al.}(2011)Cai, Song, Lam, and
  Ip}]{model-based:Cai+Song+Lam:2011}
Cai JH, Song XY, Lam KH, Ip EHS (2011).
\newblock \enquote{A Mixture of Generalized Latent Variable Models for Mixed
  Mode and Heterogeneous Data.}
\newblock \emph{Computational Statistics \& Data Analysis}, \textbf{55}(11),
  2889--2907.
\newblock \doi{10.1016/j.csda.2011.05.011}.

\bibitem[{Celeux and Govaert(1992)}]{model-based:Celeux+Govaert:1992}
Celeux G, Govaert G (1992).
\newblock \enquote{A Classification {EM} Algorithm for Clustering and two
  Stochastic Versions.}
\newblock \emph{Computational Statistics \& Data Analysis}, \textbf{14}(3),
  315--332.
\newblock \doi{10.1016/0167-9473(92)90042-e}.

\bibitem[{Celeux and Govaert(1995)}]{model-based:Celeux+Govaert:1995}
Celeux G, Govaert G (1995).
\newblock \enquote{{G}aussian Parsimonious Clustering Models.}
\newblock \emph{Pattern Recognition}, \textbf{28}(5), 781--793.
\newblock \doi{10.1016/0031-3203(94)00125-6}.

\bibitem[{Celeux \emph{et~al.}(2005)Celeux, Martin, and
  Lavergne}]{model-based:Celeux+Martin+Lavergne:2005}
Celeux G, Martin O, Lavergne C (2005).
\newblock \enquote{Mixture of Linear Mixed Models for Clustering Gene
  Expression Profiles from Repeated Microarray Experiments.}
\newblock \emph{Statistical Modelling}, \textbf{5}(3), 243--267.
\newblock \doi{10.1191/1471082x05st096oa}.

\bibitem[{Celeux and Soromenho(1996)}]{model-based:Celeux+Soromenho:1996}
Celeux G, Soromenho G (1996).
\newblock \enquote{An Entropy Criterion for Assessing the Number of Clusters in
  a Mixture Model.}
\newblock \emph{Journal of Classification}, \textbf{13}(2), 195--212.
\newblock \doi{10.1007/bf01246098}.

\bibitem[{Chan \emph{et~al.}(2008)Chan, Feng, Ottinger, Foster, West, and
  Kepler}]{model-based:Chan+Feng+Ottinger:2008}
Chan C, Feng F, Ottinger J, Foster D, West M, Kepler TB (2008).
\newblock \enquote{Statistical Mixture Modelling for Cell Subtype
  Identification in Flow Cytometry.}
\newblock \emph{Cytometry A}, \textbf{73}(8), 693--701.
\newblock \doi{10.1002/cyto.a.20583}.

\bibitem[{Chung \emph{et~al.}(2004)Chung, Loken, and
  Schafer}]{model-based:Chung+Loken:Schafer:2004}
Chung H, Loken E, Schafer JL (2004).
\newblock \enquote{Difficulties in Drawing Inferences With Finite-Mixture
  Models: A Simple Example with A Simple Solution.}
\newblock \emph{The American Statistician}, \textbf{58}(2), 152--158.
\newblock \doi{10.1198/0003130043286}.

\bibitem[{Collins and Lanza(2010)}]{model-based:Collins+Lanza:2010}
Collins LM, Lanza ST (2010).
\newblock \emph{Latent Class and Latent Transition Analysis With Applications
  in the Social, Behavioral, and Health Sciences}.
\newblock John Wiley \& Sons.

\bibitem[{Davies and Bouldin(1979)}]{model-based:Davies+Bouldin:1979}
Davies DL, Bouldin DW (1979).
\newblock \enquote{A Cluster Separation Measure.}
\newblock \emph{IEEE Transactions on Pattern Analysis and Machine
  Intelligence}, \textbf{1}(2), 224--227.
\newblock \doi{10.1109/tpami.1979.4766909}.

\bibitem[{Dayton and Macready(1988)}]{model-based:Dayton+Macready:1988}
Dayton CM, Macready GB (1988).
\newblock \enquote{Concomitant-Variable Latent-Class Models.}
\newblock \emph{Journal of the American Statistical Association},
  \textbf{83}(401), 173--178.
\newblock \doi{10.1080/01621459.1988.10478584}.

\bibitem[{Dean and Raftery(2010)}]{model-based:Dean+Raftery:2010}
Dean N, Raftery AE (2010).
\newblock \enquote{Latent Class Analysis Variable Selection.}
\newblock \emph{The Annals of the Institute of Statistical Mathematics},
  \textbf{62}(1), 11--35.
\newblock \doi{10.1007/s10463-009-0258-9}.

\bibitem[{{Di Zio} \emph{et~al.}(2007){Di Zio}, Guarnera, and
  Rocci}]{model-based:Dizio+Guarnera+Rocci:2007}
{Di Zio} M, Guarnera U, Rocci R (2007).
\newblock \enquote{A Mixture of Mixture Models for a Classification Problem:
  The Unity Measure Error.}
\newblock \emph{Computational Statistics \& Data Analysis}, \textbf{51}(5),
  2573--2585.
\newblock \doi{10.1016/j.csda.2006.01.001}.

\bibitem[{Dias \emph{et~al.}(2015)Dias, Vermunt, and
  Ramos}]{model-based:Dias+Vermunt+Ramos:2015}
Dias JG, Vermunt JK, Ramos S (2015).
\newblock \enquote{Clustering Financial Time Series: New Insights From an
  Extended Hidden {M}arkov Model.}
\newblock \emph{European Journal of Operational Research}, \textbf{243}(3),
  852--864.
\newblock \doi{10.1016/j.ejor.2014.12.041}.

\bibitem[{Dolnicar and Leisch(2010)}]{model-based:Dolnicar+Leisch:2010}
Dolnicar S, Leisch F (2010).
\newblock \enquote{Evaluation of Structure and Reproducibility of Cluster
  Solutions Using the Bootstrap.}
\newblock \emph{Marketing Letters}, \textbf{21}(1), 83--101.
\newblock \doi{10.1007/s11002-009-9083-4}.

\bibitem[{Dolnicar and Leisch(2014)}]{model-based:Dolnicar+Leisch:2014}
Dolnicar S, Leisch F (2014).
\newblock \enquote{Using Graphical Statistics to Better Understand Market
  Segmentation Solutions.}
\newblock \emph{International Journal of Market Research}, \textbf{56}(2),
  207--230.
\newblock \doi{10.2501/ijmr-2013-073}.

\bibitem[{Dunn(1974)}]{model-based:Dunn:1974}
Dunn JC (1974).
\newblock \enquote{Well Separated Clusters and Optimal Fuzzy Partitions.}
\newblock \emph{Journal of Cybernetics}, \textbf{4}(1), 95--104.
\newblock \doi{10.1080/01969727408546059}.

\bibitem[{Everitt \emph{et~al.}(2011)Everitt, Landau, Leese, and
  Stahl}]{model-based:Everitt+Landau+Leese:2011}
Everitt BS, Landau S, Leese M, Stahl D (2011).
\newblock \emph{Cluster Analysis}.
\newblock Wiley Series in Probability and Statistics, 5th edition. John Wiley
  \& Sons, Chichester.
\newblock \doi{10.1002/9780470977811}.

\bibitem[{Follmann and Lambert(1991)}]{model-based:Follmann+Lambert:1991}
Follmann DA, Lambert D (1991).
\newblock \enquote{Identifiability of Finite Mixtures of Logistic Regression
  Models.}
\newblock \emph{Journal of Statistical Planning and Inference}, \textbf{27}(3),
  375--381.
\newblock \doi{10.1016/0378-3758(91)90050-o}.

\bibitem[{Fowlkes and Mallows(1983)}]{model-based:Fowlkes+Mallows:1983}
Fowlkes EB, Mallows CL (1983).
\newblock \enquote{A Method for Comparing Two Hierarchical Clusterings.}
\newblock \emph{Journal of the American Statistical Association},
  \textbf{78}(383), 553--569.
\newblock \doi{10.1080/01621459.1983.10478008}.

\bibitem[{Fraiman \emph{et~al.}(2008)Fraiman, Justel, and
  Svarc}]{model-based:Fraiman+Justel+Svarc:2008}
Fraiman R, Justel A, Svarc M (2008).
\newblock \enquote{Selection of Variables for Cluster Analysis and
  Classification Rules.}
\newblock \emph{Journal of American Statistical Association},
  \textbf{103}(483), 1294--1303.
\newblock \doi{10.1198/016214508000000544}.

\bibitem[{Fraley and Raftery(2002)}]{model-based:Fraley+Raftery:2002}
Fraley C, Raftery AE (2002).
\newblock \enquote{Model-Based Clustering, Discriminant Analysis and Density
  Estimation.}
\newblock \emph{Journal of the American Statistical Association},
  \textbf{97}(458), 611--631.
\newblock \doi{10.1198/016214502760047131}.

\bibitem[{Franczak \emph{et~al.}(2014)Franczak, Browne, and
  McNicholas}]{model-based:Franczak+Browne+McNicholas:2014}
Franczak BC, Browne RP, McNicholas PD (2014).
\newblock \enquote{Mixtures of Shifted Asymmetric {L}aplace Distributions.}
\newblock \emph{IEEE Transactions in Pattern Analysis and Machine
  Intelligence}, \textbf{36}(6), 1149--1157.
\newblock \doi{10.1109/tpami.2013.216}.

\bibitem[{Fritsch and Ickstadt(2009)}]{model-based:Fritsch+Ickstadt:2009}
Fritsch A, Ickstadt K (2009).
\newblock \enquote{Improved Criteria for Clustering Based on the Posterior
  Similarity Matrix.}
\newblock \emph{Bayesian Analysis}, \textbf{4}(2), 367--391.
\newblock \doi{10.1214/09-ba414}.

\bibitem[{Fr{\"u}hwirth-Schnatter(2001)}]{fru:mcm}
Fr{\"u}hwirth-Schnatter S (2001).
\newblock \enquote{Markov Chain {M}onte {C}arlo Estimation of Classical and
  Dynamic Switching and Mixture Models.}
\newblock \emph{Journal of the American Statistical Association},
  \textbf{96}(453), 194--209.
\newblock \doi{10.1198/016214501750333063}.

\bibitem[{Fr{\"u}hwirth-Schnatter(2006)}]{model-based:Fruehwirth-Schnatter:2006}
Fr{\"u}hwirth-Schnatter S (2006).
\newblock \emph{Finite Mixture and {M}arkov Switching Models}.
\newblock Springer Series in Statistics. Springer-Verlag, New York.

\bibitem[{Fr{\"u}hwirth-Schnatter(2011{\natexlab{a}})}]{fru:dea}
Fr{\"u}hwirth-Schnatter S (2011{\natexlab{a}}).
\newblock \enquote{Dealing with Label Switching Under Model Uncertainty.}
\newblock In K~Mengersen, CP~Robert, D~Titterington (eds.), \emph{Mixtures:
  Estimation and Applications}, chapter~10, pp. 193--218. John Wiley \& Sons,
  Chichester.

\bibitem[{Fr{\"u}hwirth-Schnatter(2011{\natexlab{b}})}]{model-based:Fruehwirth-Schnatter:2011}
Fr{\"u}hwirth-Schnatter S (2011{\natexlab{b}}).
\newblock \enquote{Panel Data Analysis: A Survey on Model-Based Clustering of
  Time Series.}
\newblock \emph{Advances in Data Analysis and Classification}, \textbf{5}(4),
  251--280.
\newblock \doi{10.1007/s11634-011-0100-0}.

\bibitem[{Fr{\"u}hwirth-Schnatter and Kaufmann(2008)}]{fru-kau:mod}
Fr{\"u}hwirth-Schnatter S, Kaufmann S (2008).
\newblock \enquote{Model-Based Clustering of Multiple Time Series.}
\newblock \emph{Journal of Business \& Economic Statistics}, \textbf{26}(1),
  78--89.
\newblock \doi{10.1198/073500107000000106}.

\bibitem[{Fr{\"u}hwirth-Schnatter \emph{et~al.}(2012)Fr{\"u}hwirth-Schnatter,
  Pamminger, Weber, and
  Winter-Ebmer}]{model-based:Fruewirth-Schnatter+Pamminger+Weber:2012}
Fr{\"u}hwirth-Schnatter S, Pamminger C, Weber A, Winter-Ebmer R (2012).
\newblock \enquote{Labor Market Entry and Earnings Dynamics: {B}ayesian
  Inference Using Mixtures-of-Experts Markov Chain Clustering.}
\newblock \emph{Journal of Applied Econometrics}, \textbf{27}(7), 1116--1137.
\newblock \doi{10.1002/jae.1249}.

\bibitem[{Fr{\"u}hwirth-Schnatter \emph{et~al.}(2016)Fr{\"u}hwirth-Schnatter,
  Pamminger, Weber, and
  Winter-Ebmer}]{model-based:Fruewirth-Schnatter+Pamminger+Weber:2016}
Fr{\"u}hwirth-Schnatter S, Pamminger C, Weber A, Winter-Ebmer R (2016).
\newblock \enquote{Mothers' Long-Run Career Patterns After First Birth.}
\newblock \emph{Journal of the Royal Statistical Society A}, \textbf{179}(3),
  707--725.
\newblock \doi{10.1111/rssa.12151}.

\bibitem[{Fr{\"u}hwirth-Schnatter and
  Pyne(2010)}]{model-based:Fruehwirth-Schnatter+Pyne:2010}
Fr{\"u}hwirth-Schnatter S, Pyne S (2010).
\newblock \enquote{Bayesian Inference for Finite Mixtures of Univariate and
  Multivariate Skew-Normal and Skew-$t$ Distributions.}
\newblock \emph{Biostatistics}, \textbf{11}(2), 317--336.
\newblock \doi{10.1093/biostatistics/kxp062}.

\bibitem[{Fr{\"u}hwirth-Schnatter
  \emph{et~al.}(2004{\natexlab{a}})Fr{\"u}hwirth-Schnatter, T{\"u}chler, and
  Otter}]{model-based:Fruehwirth-Schnatter+Tuechler+Otter:2004}
Fr{\"u}hwirth-Schnatter S, T{\"u}chler R, Otter T (2004{\natexlab{a}}).
\newblock \enquote{Bayesian Analysis of the Heterogeneity Model.}
\newblock \emph{Journal of Business \& Economic Statistics}, \textbf{22}(1),
  2--15.
\newblock \doi{10.1198/073500103288619331}.

\bibitem[{Fr{\"u}hwirth-Schnatter
  \emph{et~al.}(2004{\natexlab{b}})Fr{\"u}hwirth-Schnatter, T{\"u}chler, and
  Otter}]{model-based:Fruehwirth-Schnatter+Tuechler+Otter:2004a}
Fr{\"u}hwirth-Schnatter S, T{\"u}chler R, Otter T (2004{\natexlab{b}}).
\newblock \enquote{Capturing Consumer Heterogeneity in Metric Conjoint Analysis
  Using {B}ayesian Mixture Models.}
\newblock \emph{International Journal of Research in Marketing},
  \textbf{21}(3), 285--297.
\newblock \doi{10.1016/j.ijresmar.2003.11.002}.

\bibitem[{Gnanadesikan \emph{et~al.}(1995)Gnanadesikan, Kettenring, and
  Tsao}]{model-based:Gnanadesikan+Kettenring+Tsao:1995}
Gnanadesikan R, Kettenring JR, Tsao SL (1995).
\newblock \enquote{Weighting and Selection of Variables for Cluster Analysis.}
\newblock \emph{Journal of Classification}, \textbf{12}(1), 113--136.
\newblock \doi{10.1007/bf01202271}.

\bibitem[{Gollini and Murphy(2014)}]{model-based:Gollini+Murphy:2014}
Gollini I, Murphy TB (2014).
\newblock \enquote{Mixture of Latent Trait Analyzers for Model-Based Clustering
  of Categorical Data.}
\newblock \emph{Statistics and Computing}, \textbf{24}(4), 569--588.
\newblock \doi{10.1007/s11222-013-9389-1}.

\bibitem[{Goodman(1974)}]{model-based:Goodman:1974}
Goodman LA (1974).
\newblock \enquote{Exploratory Latent Structure Analysis Using Both
  Identifiable and Unidentifiable Models.}
\newblock \emph{Biometrika}, \textbf{61}(2), 215--231.
\newblock \doi{10.1093/biomet/61.2.215}.

\bibitem[{Gr{\"u}n and Leisch(2008)}]{model-based:Gruen+Leisch:2008}
Gr{\"u}n B, Leisch F (2008).
\newblock \enquote{Identifiability of Finite Mixtures of Multinomial Logit
  Models with Varying and Fixed Effects.}
\newblock \emph{Journal of Classification}, \textbf{25}(2), 225--247.
\newblock \doi{10.1007/s00357-008-9022-8}.

\bibitem[{Gr{\"u}n \emph{et~al.}(2012)Gr{\"u}n, Scharl, and
  Leisch}]{model-based:Gruen+Scharl+Leisch:2012}
Gr{\"u}n B, Scharl T, Leisch F (2012).
\newblock \enquote{Modelling Time Course Gene Expression Data With Finite
  Mixtures of Linear Additive Models.}
\newblock \emph{Bioinformatics}, \textbf{28}(2), 222--228.
\newblock \doi{10.1093/bioinformatics/btr653}.

\bibitem[{Halkidi \emph{et~al.}(2001)Halkidi, Batistakis, and
  Vazirgiannis}]{model-based:Halkidi+Batistakis+Vazirgiannis:2001}
Halkidi M, Batistakis Y, Vazirgiannis M (2001).
\newblock \enquote{On Clustering Validation Techniques.}
\newblock \emph{Journal of Intelligent Information Systems}, \textbf{17}(2--3),
  107--145.
\newblock \doi{10.1023/a:1012801612483}.

\bibitem[{Handcock \emph{et~al.}(2007)Handcock, Raftery, and
  Tantrum}]{model-based:Handcock+Raftery+Tantrum:2007}
Handcock MS, Raftery AE, Tantrum JM (2007).
\newblock \enquote{Model-Based Clustering for Social Networks.}
\newblock \emph{Journal of the Royal Statistical Society A}, \textbf{170}(2),
  301--322.
\newblock \doi{10.1111/j.1467-985x.2007.00471.x}.

\bibitem[{Hennig(2000)}]{model-based:Hennig:2000}
Hennig C (2000).
\newblock \enquote{Identifiability of Models for Clusterwise Linear
  Regression.}
\newblock \emph{Journal of Classification}, \textbf{17}(2), 273--296.
\newblock \doi{10.1007/s003570000022}.

\bibitem[{Hennig(2004)}]{model-based:Hennig:2004}
Hennig C (2004).
\newblock \enquote{Asymmetric Linear Dimension Reduction for Classification.}
\newblock \emph{Journal of Computational and Graphical Statistics},
  \textbf{13}(4), 930--945.
\newblock \doi{10.1198/106186004x12740}.

\bibitem[{Hennig(2007)}]{model-based:Hennig:2007}
Hennig C (2007).
\newblock \enquote{Cluster-Wise Assessment of Cluster Stability.}
\newblock \emph{Computational Statistics \& Data Analysis}, \textbf{52}(1),
  258--271.
\newblock \doi{10.1016/j.csda.2006.11.025}.

\bibitem[{Hennig(2010)}]{model-based:Hennig:2010}
Hennig C (2010).
\newblock \enquote{Methods for Merging {G}aussian Mixture Components.}
\newblock \emph{Advances in Data Analysis and Classification}, \textbf{4}(1),
  3--34.
\newblock \doi{10.1007/s11634-010-0058-3}.

\bibitem[{Hennig(2015)}]{model-based:Hennig:2015}
Hennig C (2015).
\newblock \enquote{What Are the True Clusters?}
\newblock \emph{Pattern Recognition Letters}, \textbf{64}, 53--62.
\newblock \doi{10.1016/j.patrec.2015.04.009}.

\bibitem[{Hennig and Liao(2013)}]{model-based:Hennig+Liao:2013}
Hennig C, Liao TF (2013).
\newblock \enquote{How to Find an Appropriate Clustering for Mixed-Type
  Variables with Application to Socio-Economic Stratification.}
\newblock \emph{Journal of the Royal Statistical Society C}, \textbf{62}(3),
  309--369.
\newblock \doi{10.1111/j.1467-9876.2012.01066.x}.

\bibitem[{Hubert and Arabie(1985)}]{model-based:Hubert+Arabie:1985}
Hubert L, Arabie P (1985).
\newblock \enquote{Comparing Partitions.}
\newblock \emph{Journal of Classification}, \textbf{2}(1), 193--218.
\newblock \doi{10.1007/bf01908075}.

\bibitem[{Hunt and Jorgensen(1999)}]{model-based:Hunt+Jorgensen:1999}
Hunt L, Jorgensen M (1999).
\newblock \enquote{Mixture Model Clustering Using the {MULTIMIX} Program.}
\newblock \emph{Australian \& New Zealand Journal of Statistics},
  \textbf{41}(2), 153--171.
\newblock \doi{10.1111/1467-842x.00071}.

\bibitem[{Jaccard(1912)}]{model-based:Jaccard:1912}
Jaccard P (1912).
\newblock \enquote{The Distribution of the Flora in the Alpine Zone.}
\newblock \emph{The New Phytologist}, \textbf{11}(2), 37--50.
\newblock \doi{10.1111/j.1469-8137.1912.tb05611.x}.

\bibitem[{Jasra \emph{et~al.}(2005)Jasra, Holmes, and
  Stephens}]{model-based:Jasra+Holmes+Stephens:2005}
Jasra A, Holmes CC, Stephens DA (2005).
\newblock \enquote{Markov Chain {M}onte {C}arlo Methods and the Label Switching
  Problem in {B}ayesian Mixture Modelling.}
\newblock \emph{Statistical Science}, \textbf{20}(1), 50--67.
\newblock \doi{10.1214/088342305000000016}.

\bibitem[{Kaufman and Rousseeuw(1990)}]{model-based:Kaufman+Rousseeuw:1990}
Kaufman L, Rousseeuw PJ (1990).
\newblock \emph{Finding Groups in Data: An Introduction to Cluster Analysis}.
\newblock John Wiley \& Sons, New York, USA.
\newblock \doi{10.1002/9780470316801}.

\bibitem[{Kebschull \emph{et~al.}(2014)Kebschull, Demmer, Gr{\"u}n, Guarnieri,
  Pavlidis, and Papapanou}]{model-based:Kebschull+Demmer+Gruen:2014}
Kebschull M, Demmer RT, Gr{\"u}n B, Guarnieri P, Pavlidis P, Papapanou PN
  (2014).
\newblock \enquote{Gingival Tissue Transcriptomes Identify Distinct
  Periodontitis Phenotypes.}
\newblock \emph{Journal of Dental Research}, \textbf{93}(5), 459--468.
\newblock \doi{10.1177/0022034514527288}.

\bibitem[{Keribin(2000)}]{model-based:Keribin:2000}
Keribin C (2000).
\newblock \enquote{Consistent Estimation of the Order of Mixture Models.}
\newblock \emph{Sankhy{\=a}: The Indian Journal of Statistics A},
  \textbf{62}(1), 49--66.

\bibitem[{Kim \emph{et~al.}(2014)Kim, Yun, Park, Joo, and
  Kim}]{model-based:Kim+Yun+Park:2014}
Kim KH, Yun ST, Park SS, Joo Y, Kim TS (2014).
\newblock \enquote{Model-Based Clustering of Hydrochemical Data to Demarcate
  Natural Versus Human Impacts on Bedrock Groundwater Quality in Rural Areas,
  {S}outh {K}orea.}
\newblock \emph{Journal of Hydrology}, \textbf{519}(Part A), 626--636.
\newblock \doi{10.1016/j.jhydrol.2014.07.055}.

\bibitem[{Kim \emph{et~al.}(2006)Kim, Tadesse, and
  Vannucci}]{model-based:Kim+Tadesse+Vannucci:2006}
Kim S, Tadesse MG, Vannucci M (2006).
\newblock \enquote{Variable Selection in Clustering via {D}irichlet Process
  Mixture Models.}
\newblock \emph{Biometrika}, \textbf{93}(4), 877--893.
\newblock \doi{10.1093/biomet/93.4.877}.

\bibitem[{Lau and Green(2007)}]{model-based:Lau+Green:2007}
Lau JW, Green P (2007).
\newblock \enquote{Bayesian Model-Based Clustering Procedures.}
\newblock \emph{Journal of Computational and Graphical Statistics},
  \textbf{16}(3), 526--558.
\newblock \doi{10.1198/106186007x238855}.

\bibitem[{Lee and McLachlan(2013)}]{model-based:Lee+McLachlan:2013}
Lee S, McLachlan GJ (2013).
\newblock \enquote{Model-Based Clustering and Classification With Non-Normal
  Mixture Distributions.}
\newblock \emph{Statistical Methods and Applications}, \textbf{22}(4),
  427--454.
\newblock \doi{10.1007/s10260-013-0237-4}.

\bibitem[{Lee and McLachlan(2014)}]{model-based:Lee+McLachlan:2014}
Lee S, McLachlan GJ (2014).
\newblock \enquote{Finite Mixtures of Multivariate Skew $t$-Distributions: Some
  Recent and New Results.}
\newblock \emph{Statistics and Computing}, \textbf{24}(2), 181--202.
\newblock \doi{10.1007/s11222-012-9362-4}.

\bibitem[{Leisch(2006)}]{model-based:Leisch:2006}
Leisch F (2006).
\newblock \enquote{A Toolbox for $K$-Centroids Cluster Analysis.}
\newblock \emph{Computational Statistics \& Data Analysis}, \textbf{51}(2),
  526--544.
\newblock \doi{10.1016/j.csda.2005.10.006}.

\bibitem[{Leisch(2010)}]{model-based:Leisch:2010}
Leisch F (2010).
\newblock \enquote{Neighborhood Graphis, Stripes, and Shadow Plots for Cluster
  Visualization.}
\newblock \emph{Statistics and Computing}, \textbf{20}(4), 457--469.
\newblock \doi{10.1007/s11222-009-9137-8}.

\bibitem[{Li(2005)}]{model-based:Li:2005}
Li J (2005).
\newblock \enquote{Clustering Based on a Multilayer Mixture Model.}
\newblock \emph{Journal of Computational and Graphical Statistics},
  \textbf{3}(14), 547--568.
\newblock \doi{10.1198/106186005x59586}.

\bibitem[{Linzer and Lewis(2011)}]{model-based:Linzer+Lewis:2011}
Linzer D, Lewis J (2011).
\newblock \enquote{{poLCA}: An {R} Package for Polytomous Variable Latent Class
  Analysis.}
\newblock \emph{Journal of Statistical Software, Articles}, \textbf{42}(10),
  1--29.
\newblock \doi{10.18637/jss.v042.i10}.

\bibitem[{Luan and Li(2003)}]{model-based:Luan+Li:2003}
Luan Y, Li H (2003).
\newblock \enquote{Clustering of Time-Course Gene Expression Data Using a
  Mixed-Effects Model with {B}-Splines.}
\newblock \emph{Bioinformatics}, \textbf{19}(4), 474--482.
\newblock \doi{10.1093/bioinformatics/btg014}.

\bibitem[{Luan and Li(2004)}]{model-based:Luan+Li:2004}
Luan Y, Li H (2004).
\newblock \enquote{Model-Based Methods for Identifying Periodically Expressed
  Genes Based on Time Course Microarray Gene Expression Data.}
\newblock \emph{Bioinformatics}, \textbf{20}(3), 332--339.
\newblock \doi{10.1093/bioinformatics/btg413}.

\bibitem[{Malsiner-Walli \emph{et~al.}(2016)Malsiner-Walli,
  Fr{\"u}hwirth-Schnatter, and
  Gr{\"u}n}]{model-based:Malsiner-Walli+Fruehwirth-Schnatter+Gruen:2016}
Malsiner-Walli G, Fr{\"u}hwirth-Schnatter S, Gr{\"u}n B (2016).
\newblock \enquote{Model-Based Clustering Based on Sparse Finite {G}aussian
  Mixtures.}
\newblock \emph{Statistics and Computing}, \textbf{26}(1), 303--324.
\newblock \doi{10.1007/s11222-014-9500-2}.

\bibitem[{Malsiner-Walli \emph{et~al.}(2017)Malsiner-Walli,
  Fr{\"u}hwirth-Schnatter, and
  Gr{\"u}n}]{model-based:Malsiner-Walli+Fruehwirth-Schnatter+Gruen:2017}
Malsiner-Walli G, Fr{\"u}hwirth-Schnatter S, Gr{\"u}n B (2017).
\newblock \enquote{Identifying Mixtures of Mixtures Using {B}ayesian
  Estimation.}
\newblock \emph{Journal of Computational and Graphical Statistics},
  \textbf{26}(2), 285--295.
\newblock \doi{10.1080/10618600.2016.1200472}.

\bibitem[{Marin \emph{et~al.}(2005)Marin, Mengersen, and
  Robert}]{model-based:Marin+Mengersen+Robert:2005}
Marin JM, Mengersen K, Robert CP (2005).
\newblock \enquote{Bayesian Modelling and Inference on Mixtures of
  Distributions.}
\newblock In D~Dey, CR~Rao (eds.), \emph{Bayesian Thinking: Modeling and
  Computation}, volume~25 of \emph{Handbook of Statistics}, chapter~16, pp.
  459--507. North--Holland, Amsterdam.

\bibitem[{Maugis \emph{et~al.}(2009)Maugis, Celeux, and
  Martin-Magniette}]{model-based:Maugis+Celeux+Martin:2009}
Maugis C, Celeux G, Martin-Magniette ML (2009).
\newblock \enquote{Variable Selection for Clustering with {G}aussian Mixture
  Models.}
\newblock \emph{Biometrics}, \textbf{65}(3), 701--709.
\newblock \doi{10.1111/j.1541-0420.2008.01160.x}.

\bibitem[{McLachlan(1982)}]{model-based:McLachlan:1982}
McLachlan GJ (1982).
\newblock \enquote{The Classification and Mixture Maximum Likelihood Approaches
  to Cluster Analysis.}
\newblock In PR~Krishnaiah, LN~Kanal (eds.), \emph{Handbook of Statistics:
  Classification Pattern Recognition and Reduction of Dimensionality},
  volume~2, pp. 199--208. Elsevier.

\bibitem[{McLachlan \emph{et~al.}(2004)McLachlan, Do, and
  Ambroise}]{model-based:McLachlan+Do+Ambroise:2004}
McLachlan GJ, Do KA, Ambroise C (2004).
\newblock \emph{Analyzing Microarray Gene Expression Data}.
\newblock Wiley Series in Probability and Statistics. John Wiley \& Sons.
\newblock \doi{10.1002/047172842x}.

\bibitem[{McLachlan and
  Peel(2000{\natexlab{a}})}]{model-based:McLachlan+Peel:2000b}
McLachlan GJ, Peel D (2000{\natexlab{a}}).
\newblock \emph{Finite Mixture Models}.
\newblock John Wiley \& Sons.
\newblock \doi{10.1002/0471721182}.

\bibitem[{McLachlan and
  Peel(2000{\natexlab{b}})}]{model-based:McLachlan+Peel:2000}
McLachlan GJ, Peel D (2000{\natexlab{b}}).
\newblock \enquote{Mixtures of Factor Analyzers.}
\newblock In \emph{Proceedings of the Seventeenth International Conference on
  Machine Learning}, pp. 599--606. Morgan Kaufmann, San Francisco.

\bibitem[{McLachlan \emph{et~al.}(2003)McLachlan, Peel, and
  Bean}]{model-based:McLachlan+Peel+Bean:2003}
McLachlan GJ, Peel D, Bean RW (2003).
\newblock \enquote{Modelling High-Dimensional Data by Mixtures of Factor
  Analyzers.}
\newblock \emph{Computational Statistics \& Data Analysis}, \textbf{41}(3--4),
  379--388.
\newblock \doi{10.1016/s0167-9473(02)00183-4}.

\bibitem[{McNicholas(2016{\natexlab{a}})}]{model-based:McNicholas:2016a}
McNicholas PD (2016{\natexlab{a}}).
\newblock \emph{Mixture Model-Based Classification}.
\newblock CRC Press.

\bibitem[{McNicholas(2016{\natexlab{b}})}]{model-based:McNicholas:2016}
McNicholas PD (2016{\natexlab{b}}).
\newblock \enquote{Model-Based Clustering.}
\newblock \emph{Journal of Classification}, \textbf{33}(3), 331--373.
\newblock \doi{10.1007/s00357-016-9211-9}.

\bibitem[{McNicholas and Murphy(2008)}]{model-based:McNicholas+Murphy:2008}
McNicholas PD, Murphy TB (2008).
\newblock \enquote{Parsimonious {G}aussian Mixture Models.}
\newblock \emph{Statistics and Computing}, \textbf{18}(3), 285--296.
\newblock \doi{10.1007/s11222-008-9056-0}.

\bibitem[{McNicholas and Murphy(2010)}]{model-based:McNicholas+Murphy:2010}
McNicholas PD, Murphy TB (2010).
\newblock \enquote{Model-Based Clustering of Microarray Expression Data via
  Latent {G}aussian Mixture Models.}
\newblock \emph{Bioinformatics}, \textbf{26}(21), 2705--2712.
\newblock \doi{10.1093/bioinformatics/btq498}.

\bibitem[{Melnykov(2016)}]{model-based:Melnykov:2016}
Melnykov V (2016).
\newblock \enquote{Merging Mixture Components for Clustering Through Pairwise
  Overlap.}
\newblock \emph{Journal of Computational and Graphical Statistics},
  \textbf{25}(1), 66--90.
\newblock \doi{10.1080/10618600.2014.978007}.

\bibitem[{Molitor \emph{et~al.}(2010)Molitor, Papathomas, Jerrett, and
  Richardson}]{model-based:Molitor+Papathomas+Jerrett:2010}
Molitor J, Papathomas M, Jerrett M, Richardson S (2010).
\newblock \enquote{Bayesian Profile Regression with an Application to the
  {N}ational Survey of Children's Health.}
\newblock \emph{Biostatistics}, \textbf{11}(3), 484--498.
\newblock \doi{10.1093/biostatistics/kxq013}.

\bibitem[{Muth{\'e}n and Asparouhov(2015)}]{model-based:Muthen+Asparouhov:2015}
Muth{\'e}n B, Asparouhov T (2015).
\newblock \enquote{Growth Mixture Modeling with Non-Normal Distributions.}
\newblock \emph{Statistics in Medicine}, \textbf{34}(6), 1041--1058.
\newblock \doi{10.1002/sim.6388}.

\bibitem[{Muth{\'e}n \emph{et~al.}(2002)Muth{\'e}n, Brown, Masyn, Jo, Khoo,
  Yang, Wang, Kellam, Carlin, and Liao}]{model-based:Muthen+Brown+Masyn:2002}
Muth{\'e}n B, Brown CH, Masyn K, Jo B, Khoo ST, Yang CC, Wang CP, Kellam SG,
  Carlin JB, Liao J (2002).
\newblock \enquote{General Growth Mixture Modeling for Randomized Preventive
  Interventions.}
\newblock \emph{Biostatistics}, \textbf{3}(4), 459--475.
\newblock \doi{10.1093/biostatistics/3.4.459}.

\bibitem[{Newman and Leicht(2007)}]{model-based:Newman+Leicht:2007}
Newman MEJ, Leicht EA (2007).
\newblock \enquote{Mixture Models and Exploratory Analysis in Networks.}
\newblock \emph{Proceedings of the National Academy of Sciences of the United
  States of America}, \textbf{104}(23), 9564--9569.
\newblock \doi{10.1073/pnas.0610537104}.

\bibitem[{Ng \emph{et~al.}(2006)Ng, McLachlan, Wang, Jones, and
  Ng}]{model-based:Ng+McLachlan+Wang:2006}
Ng SK, McLachlan GJ, Wang K, Jones LBT, Ng SW (2006).
\newblock \enquote{A Mixture Model with Random-Effects Components for
  Clustering Correlated Gene-Expression Profiles.}
\newblock \emph{Bioinformatics}, \textbf{22}(14), 1745--1752.
\newblock \doi{10.1093/bioinformatics/btl165}.

\bibitem[{O'Hagan \emph{et~al.}(2016)O'Hagan, Murphy, Gormley, McNicholas, and
  Karlis}]{model-based:Hagan+Murphy+Gormley:2016}
O'Hagan A, Murphy TB, Gormley IC, McNicholas PD, Karlis D (2016).
\newblock \enquote{Clustering with the Multivariate Normal Inverse {G}aussian
  Distribution.}
\newblock \emph{Computational Statistics \& Data Analysis}, \textbf{93},
  18--30.
\newblock \doi{10.1016/j.csda.2014.09.006}.

\bibitem[{Ohlssen \emph{et~al.}(2006)Ohlssen, Sharples, and
  Spiegelhalter}]{model-based:Ohlssen+Sharples+Spiegelhalter:2006}
Ohlssen DI, Sharples LD, Spiegelhalter DJ (2006).
\newblock \enquote{Flexible Random-Effects Models Using {B}ayesian
  Semi-Parametric Models: Applications to Institutional Comparisons.}
\newblock \emph{Statistics in Medicine}, \textbf{26}(9), 2088--2112.
\newblock \doi{10.1002/sim.2666}.

\bibitem[{Pan and Shen(2007)}]{model-based:Pan+Shen:2007}
Pan W, Shen X (2007).
\newblock \enquote{Penalized Model-Based Clustering with Application to
  Variable Selection.}
\newblock \emph{Journal of Machine Learning Research}, \textbf{8}, 1145--1164.

\bibitem[{Papastamoulis(2016)}]{model-based:Papastamoulis:2016}
Papastamoulis P (2016).
\newblock \enquote{{label.switching}: An {R} Package for Dealing with the Label
  Switching Problem in {MCMC} Outputs.}
\newblock \emph{Journal of Statistical Software}, \textbf{69}(1), 1--24.
\newblock \doi{10.18637/jss.v069.c01}.

\bibitem[{Perera and Mo(2016)}]{model-based:Perera+Mo:2016}
Perera LP, Mo B (2016).
\newblock \enquote{Data Analysis on Marine Engine Operating Regions In Relation
  to Ship Navigation.}
\newblock \emph{Ocean Engineering}, \textbf{128}, 163--172.
\newblock \doi{10.1016/j.oceaneng.2016.10.029}.

\bibitem[{Pyne \emph{et~al.}(2009)Pyne, Hu, Wang, Rossin, Lin, Maier,
  Baecher-Allan, McLachlan, Tamayo, Hafler, {De Jager}, and
  Mesirov}]{model-based:Pyne+Hu+Wang:2009}
Pyne S, Hu X, Wang K, Rossin E, Lin TI, Maier LM, Baecher-Allan C, McLachlan
  GJ, Tamayo P, Hafler DA, {De Jager} PL, Mesirov JP (2009).
\newblock \enquote{Automated High-Dimensional Flow Cytometric Data Analysis.}
\newblock \emph{Proceedings of the National Academy of Sciences},
  \textbf{106}(21), 8519--8524.
\newblock \doi{10.1073/pnas.0903028106}.

\bibitem[{Raftery and Dean(2006)}]{model-based:Raftery+Dean:2006}
Raftery AE, Dean N (2006).
\newblock \enquote{Variable Selection for Model-Based Clustering.}
\newblock \emph{Journal of the American Statistical Association},
  \textbf{101}(473), 168--178.
\newblock \doi{10.1198/016214506000000113}.

\bibitem[{Rand(1971)}]{model-based:Rand:1971}
Rand WM (1971).
\newblock \enquote{Objective Criteria for the Evaluation of Clustering
  Methods.}
\newblock \emph{Journal of the American Statistical Association},
  \textbf{66}(336), 846--850.
\newblock \doi{10.1080/01621459.1971.10482356}.

\bibitem[{Redner and Walker(1984)}]{model-based:Redner+Walker:1984}
Redner RA, Walker HF (1984).
\newblock \enquote{Mixture Densities, Maximum Likelihood and the {EM}
  Algorithm.}
\newblock \emph{SIAM Review}, \textbf{26}(2), 195--239.
\newblock \doi{10.1137/1026034}.

\bibitem[{Rosen and Tanner(1999)}]{model-based:Rosen+Tanner:1999}
Rosen O, Tanner M (1999).
\newblock \enquote{Mixtures of Proportional Hazards Regression Models.}
\newblock \emph{Statistics in Medicine}, \textbf{18}(9), 1119--1131.
\newblock
  \doi{10.1002/(sici)1097-0258(19990515)18:9<1119::aid-sim116>3.0.co;2-v}.

\bibitem[{Rousseeuw(1987)}]{model-based:Rousseeuw:1987}
Rousseeuw PJ (1987).
\newblock \enquote{Silhouettes: A Graphical Aid to the Interpretation and
  Validation of Cluster Analysis.}
\newblock \emph{Computational and Applied Mathematics}, \textbf{20}, 53--65.
\newblock \doi{10.1016/0377-0427(87)90125-7}.

\bibitem[{Scharl \emph{et~al.}(2010)Scharl, Gr{\"u}n, and
  Leisch}]{model-based:Scharl+Gruen+Leisch:2010}
Scharl T, Gr{\"u}n B, Leisch F (2010).
\newblock \enquote{Mixtures of Regression Models for Time-Course Gene
  Expression Data: Evaluation of Initialization and Random Effects.}
\newblock \emph{Bioinformatics}, \textbf{26}(3), 370--377.
\newblock \doi{10.1093/bioinformatics/btp686}.

\bibitem[{Scott and Symons(1971)}]{model-based:Scott+Symons:1971}
Scott AJ, Symons MJ (1971).
\newblock \enquote{Clustering Methods Based on Likelihood Ratio Criteria.}
\newblock \emph{Biometrics}, \textbf{27}(2), 387--397.
\newblock \doi{10.2307/2529003}.

\bibitem[{Scrucca(2010)}]{model-based:Scrucca:2010}
Scrucca L (2010).
\newblock \enquote{Dimension Reduction for Model-Based Clustering.}
\newblock \emph{Statistics and Computing}, \textbf{20}(4), 471--484.
\newblock \doi{10.1007/s11222-009-9138-7}.

\bibitem[{Scrucca(2016)}]{model-based:Scrucca:2016}
Scrucca L (2016).
\newblock \enquote{Identifying Connected Components in {G}aussian Finite
  Mixture Models for Clustering.}
\newblock \emph{Computational Statistics \& Data Analysis}, \textbf{93}, 5--17.
\newblock \doi{10.1016/j.csda.2015.01.006}.

\bibitem[{Skakun \emph{et~al.}(2017)Skakun, Franch, Vermote, Roger,
  Becker-Reshef, Justice, and Kussul}]{model-based:Skakun+Franch+Vermote:2017}
Skakun S, Franch B, Vermote E, Roger JC, Becker-Reshef I, Justice C, Kussul N
  (2017).
\newblock \enquote{Early Season Large-Area Winter Crop Mapping Using {MODIS}
  {NDVI} Data, Growing Degree Days Information and a {G}aussian Mixture Model.}
\newblock \emph{Remote Sensing of Environment}, \textbf{195}, 244--258.
\newblock \doi{10.1016/j.rse.2017.04.026}.

\bibitem[{Sp{\"a}th(1979)}]{model-based:Spaeth:1979}
Sp{\"a}th H (1979).
\newblock \enquote{Algorithm 39 {C}lusterwise Linear Regression.}
\newblock \emph{Computing}, \textbf{22}(4), 367--373.
\newblock \doi{10.1007/bf02265317}.

\bibitem[{Sperrin \emph{et~al.}(2010)Sperrin, Jaki, and
  Wit}]{model-based:Sperrin+Jaki+Wit:2010}
Sperrin M, Jaki T, Wit E (2010).
\newblock \enquote{Probabilistic Relabelling Strategies for the Label Switching
  Problem in {B}ayesian Mixture Models.}
\newblock \emph{Statistics and Computing}, \textbf{20}(3), 357--366.
\newblock \doi{10.1007/s11222-009-9129-8}.

\bibitem[{Stahl and Sallis(2012)}]{model-based:Stahl+Sallis:2012}
Stahl D, Sallis H (2012).
\newblock \enquote{Model-Based Cluster Analysis.}
\newblock \emph{Wiley Interdisciplinary Reviews: Computational Statistics},
  \textbf{4}(4), 341--358.
\newblock \doi{10.1002/wics.1204}.

\bibitem[{Steinley and
  Brusco(2008{\natexlab{a}})}]{model-based:Steinley+Brusco:2008a}
Steinley D, Brusco MJ (2008{\natexlab{a}}).
\newblock \enquote{A New Variable Weighting and Selection Procedure for
  $K$-Means Cluster Analysis.}
\newblock \emph{Multivariate Behavioral Research}, \textbf{43}(1), 77--108.
\newblock \doi{10.1080/00273170701836695}.

\bibitem[{Steinley and
  Brusco(2008{\natexlab{b}})}]{model-based:Steinley+Brusco:2008}
Steinley D, Brusco MJ (2008{\natexlab{b}}).
\newblock \enquote{Selection of Variables in Cluster Analysis: An Empirical
  Comparison of Eight Procedures.}
\newblock \emph{Psychometrika}, \textbf{73}(1), 125--144.
\newblock \doi{10.1007/s11336-007-9019-y}.

\bibitem[{Stephens(2000)}]{model-based:Stephens:2000}
Stephens M (2000).
\newblock \enquote{Dealing with Label Switching in Mixture Models.}
\newblock \emph{Journal of the Royal Statistical Society B}, \textbf{62}(4),
  795--809.
\newblock \doi{10.1111/1467-9868.00265}.

\bibitem[{Symons(1981)}]{model-based:Symons:1981}
Symons MJ (1981).
\newblock \enquote{Clustering Criteria and Multivariate Normal Mixtures.}
\newblock \emph{Biometrics}, \textbf{37}(1), 35--43.
\newblock \doi{10.2307/2530520}.

\bibitem[{Tadesse \emph{et~al.}(2005)Tadesse, Sha, and
  Vanucci}]{model-based:Tadesse+Sha+Vanucci:2005}
Tadesse MG, Sha N, Vanucci M (2005).
\newblock \enquote{Bayesian Variable Selection in Clustering High-Dimensional
  Data.}
\newblock \emph{Journal of the American Statistical Association},
  \textbf{100}(470), 602--617.
\newblock \doi{10.1198/016214504000001565}.

\bibitem[{Tibshirani(1996)}]{model-based:Tibshirani:1996}
Tibshirani R (1996).
\newblock \enquote{Regression Shrinkage and Selection via the Lasso.}
\newblock \emph{Journal of the Royal Statistical Society B}, \textbf{58}(1),
  267--288.
\newblock \doi{10.1111/j.1467-9868.2011.00771.x}.

\bibitem[{Vrbik and McNicholas(2014)}]{model-based:Vrbik+McNicholas:2012}
Vrbik I, McNicholas PD (2014).
\newblock \enquote{Parsimonious Skew Mixture Models for Model-Based Clustering
  and Classification.}
\newblock \emph{Computational Statistics \& Data Analysis}, \textbf{71},
  196--210.
\newblock \doi{10.1016/j.csda.2013.07.008}.

\bibitem[{Wade and Gharhamani({in press})}]{model-based:Wade+Ghahramani:2018}
Wade S, Gharhamani Z ({in press}).
\newblock \enquote{Bayesian Cluster Analysis: Point Estimation and Credible
  Balls.}
\newblock \emph{Bayesian Analysis}.
\newblock \doi{10.1214/17-ba1073}.

\bibitem[{Wedel and DeSarbo(1995)}]{model-based:Wedel+DeSarbo:1995}
Wedel M, DeSarbo WS (1995).
\newblock \enquote{A Mixture Likelihood Approach for Generalized Linear
  Models.}
\newblock \emph{Journal of Classification}, \textbf{12}(1), 21--55.
\newblock \doi{10.1007/bf01202266}.

\bibitem[{Wedel and DeSarbo(2002)}]{model-based:Wedel+DeSarbo:2002}
Wedel M, DeSarbo WS (2002).
\newblock \enquote{Market Segment Derivation and Profiling via a Finite Mixture
  Model Framework.}
\newblock \emph{Marketing Letters}, \textbf{13}(1), 17--25.
\newblock \doi{10.1023/a:1015059024154}.

\bibitem[{Wedel and Kamakura(2001)}]{model-based:Wedel+Kamakura:2001}
Wedel M, Kamakura WA (2001).
\newblock \emph{Market Segmentation -- Conceptual and Methodological
  Foundations}.
\newblock 2nd edition. Kluwer Academic Publishers.

\bibitem[{White \emph{et~al.}(2016)White, Wyse, and
  Murphy}]{model-based:White+Wyse+Murphy:2016}
White A, Wyse J, Murphy TB (2016).
\newblock \enquote{Bayesian Variable Selection for Latent Class Analysis Using
  a Collapsed {G}ibbs Sampler.}
\newblock \emph{Statistics and Computing}, \textbf{26}(1), 511--527.
\newblock \doi{10.1007/s11222-014-9542-5}.

\bibitem[{Yao \emph{et~al.}(2011)Yao, Fu, and
  Lee}]{model-based:Yao+Fu+Lee:2011}
Yao F, Fu Y, Lee TCM (2011).
\newblock \enquote{Functional Mixture Regression.}
\newblock \emph{Biostatistics}, \textbf{12}(2), 341--353.
\newblock \doi{10.1093/biostatistics/kxq067}.

\bibitem[{Yau and Holmes(2011)}]{model-based:Yau+Holmes:2011}
Yau C, Holmes C (2011).
\newblock \enquote{Hierarchical {B}ayesian Nonparametric Mixture Models for
  Clustering with Variable Relevance Determination.}
\newblock \emph{Bayesian Analysis}, \textbf{6}(2), 329--352.
\newblock \doi{10.1214/11-ba612}.

\bibitem[{Yerebakan \emph{et~al.}(2014)Yerebakan, Rajwa, and
  Dundar}]{model-based:Yerebakan+Rajwa+Dundar:2014}
Yerebakan HZ, Rajwa B, Dundar M (2014).
\newblock \enquote{The Infinite Mixture of Infinite {G}aussian Mixtures.}
\newblock In \emph{Advances in Neural Information Processing Systems}, pp.
  28--36.

\bibitem[{Zhao \emph{et~al.}(2015)Zhao, Shi, Shearon, and
  Li}]{model-based:Zhao+Shi+Shearon:2015}
Zhao L, Shi J, Shearon TH, Li Y (2015).
\newblock \enquote{A {D}irichlet Process Mixture Model for Survival Outcome
  Data: Assessing Nationwide Kidney Transplant Centers.}
\newblock \emph{Statistics in Medicine}, \textbf{34}(8), 1404--1416.
\newblock \doi{10.1002/sim.6438}.

\bibitem[{Zhong and Ghosh(2003)}]{model-based:Zhong+Ghosh:2003}
Zhong S, Ghosh J (2003).
\newblock \enquote{A Unified Framework for Model-Based Clustering.}
\newblock \emph{Journal of Machine Learning Research}, \textbf{4}, 1001--1037.

\end{thebibliography}
\end{document}